\ifodd 0
	\documentclass[aps,prd,preprint,endfloats*]{revtex4-1} \usepackage{amsmath,amssymb,bm,color,graphicx,multirow,mfirstuc} \usepackage{wasysym} \usepackage{tgtermes} 
   	\usepackage{parskip}
\newcommand{\note}[1]{\textcolor{red}{#1}\ignorespacesafterend}
\newenvironment{notered}{\color{red}} {\ignorespacesafterend}
\renewcommand{\paragraph}[1]{\textcolor{red}{[#1]}}

\else
\documentclass[aps,prd,preprint,tightenlines]{revtex4} \usepackage{amsmath,amssymb,bm,color,graphicx}

	\newcommand{\note}[1]{}  \renewcommand{\paragraph}[1]{}
\fi

\usepackage[colorlinks=true,citecolor=blue,linkcolor=blue,urlcolor=blue]{hyperref}

\usepackage{comment}

\newcommand{\Sec}[1]{Sec.~\ref{#1}}

\newcommand{\fig}[1]{Fig.~\ref{#1}}

\newcommand{\tab}[1]{Tab.~\ref{#1}}

\newcommand{\eq}[1]{Eq.~(\ref{#1})}

\newcommand{\galprop}{\texttt{GalProp}}
\newcommand{\galpropv}{\texttt{GalProp-v54}}
\newcommand{\helmod}{\texttt{HelMod}}
\newcommand{\helmodv}{\texttt{HelMod-v4.1}}
\newcommand{\darksusy}{\texttt{DarkSUSY-v6.2}}
\newcommand{\damatis}{\texttt{DAMATIS}}
\newcommand{\damascus}{\texttt{DaMaSCUS}}
\newcommand{\damascuscrust}{\texttt{DaMaSCUS-CRUST}}
\newcommand{\earthshadow}{\texttt{EarthShadow}}
\newcommand{\verne}{\texttt{VERNE}}
\newcommand{\realearthscatterdm}{\texttt{realEarthScatterDM}}
\newcommand{\darkprop}{\texttt{DarkProp}}

\begin{document} 
\title{Production and attenuation of cosmic-ray boosted dark matter}

\author{Chen Xia$^a$}\author{Yan-Hao Xu$^a$}\author{Yu-Feng Zhou$^{a,b,c}$}\affiliation{
  $^a$CAS key laboratory of theoretical Physics, Institute of
  Theoretical Physics, Chinese Academy of Sciences, Beijing 100190,
  China; \\ School of Physics, University of Chinese Academy of
  Sciences, Beijing 100049, China;\\ $^b$School of Fundamental Physics
  and Mathematical Sciences, Hangzhou Institute for Advanced Study,
  UCAS, Hangzhou 310024, China;\\ $^c$International Centre for
  Theoretical Physics Asia-Pacific, Beijing/Hangzhou, China.}

\begin{abstract}
Light sub-GeV halo dark matter (DM) particles up-scattered by
high-energy cosmic-rays (CRs) (referred to as CRDM) can be energetic
and become detectable by conventional DM direct detection
experiments. We perform a refined analysis on the exclusion bounds of
the spin-independent DM-nucleon scattering cross section $\sigma_{\chi
p}$ in this approach. For the exclusion lower bounds, we determine the
parameter of the effective distance $D_\text{eff}$ for CRDM production
using spatial-dependent CR fluxes and including the contributions from
the major heavy CR nuclear species. We obtain $D_\text{eff}\simeq
9$~kpc for CRDM particles with kinetic energy above $\sim
1~\text{GeV}$, which pushes the corresponding exclusion lower bounds
down to $\sigma_{\chi p} \sim 4\times 10^{-32}~\text{cm}^2$ for DM
particle mass at MeV scale and below.  For the exclusion upper bounds
from Earth attenuation, previous estimations neglecting the nuclear
form factor leaded to typical exclusion upper bounds of $\sigma_{\chi
p}\sim\mathcal{O}(10^{-28})~\text{cm}^2$ from the XENON1T data. Using
both the analytic and numerical approaches, we show that for CRDM
particles, the presence of the nuclear form factor strongly suppresses
the effect of Earth attenuation. Consequently, the cross section that
can be excluded by the XENON1T data can be a few orders of magnitude
higher, which closes the gap in the cross sections excluded by the
XENON1T experiment and that by the astrophysical measurements such
that for the cosmic microwave background (CMB), galactic gas cloud
cooling, and structure formation, etc..
\end{abstract}
\date{\today}
\maketitle

\section{Introduction}\label{sec:introduction}
Although compelling astrophysical evidence supports the existence of
dark matter (DM) in the Universe, whether or not DM participates
non-gravitational interactions remains unknown. Underground DM direct
detection (DD) experiments search for recoil signals from the possible
scatterings between the halo DM particles and target nuclei.
Stringent constraints on the DM-nucleon scattering cross section have
been established, which reaches $\sigma_{\chi
p}\leq \mathcal{O}(10^{-46})~\text{cm}^2$ for DM particle mass around
$\mathcal{O}(10)~\text{GeV}$~\cite{Aprile:2018dbl,PandaX-II:2016vec}. As
the typical detection threshold of the current experiments is of
$\mathcal{O}(\text{keV})$, for the standard halo DM with a typical
escape velocity of $\sim540~\text{km s}^{-1}$, the current DD
experiments lose sensitivity rapidly to DM particles with mass below a
few GeV. The reason is that for lighter halo DM particles the kinetic
energy is lower, and the energy transfer to the target nuclei is less
efficient. Several physical processes have been considered to lower
the detection threshold such as using additional photon emission in
the inelastic scattering process~\cite{Kouvaris:2016afs} and the
Migdal effect~\cite{Ibe:2017yqa,Dolan:2017xbu}, etc..

\paragraph{Astro.}
The same DM-nucleus scattering process can leave imprints in some
astrophysical observables, which can be used to place constraints on
the scattering cross section. The resulting constraints can be applied
to much lower DM particle masses unreachable to the current DM direct
detection experiments, but the constraints are in general much weaker.
For instance, from the spectral distortion of the cosmic microwave
background (CMB), a constraint of $\sigma_{\chi
p}\lesssim5\times10^{-27}~\text{cm}^{2}$ for $m_{\chi}$ in the range
of 1~keV-TeV can be obtained~\cite{Gluscevic:2017ywp}.  The
measurement of the gas cooling rate of the Leo T dwarf galaxy can also
lead to a constraint of $\sigma_{\chi
p}\lesssim3\times10^{-25}~\text{cm}^{2}$ for $m_{\chi}\lesssim$
1~MeV~\cite{Wadekar:2019xnf,Bhoonah:2018wmw}.  The constraints from
the population of Milky Way (MW) satellite galaxies can reach
$\sigma_{\chi p}\lesssim6\times10^{-30}~\text{cm}^{2}$ for
$m_{\chi}\lesssim10~\text{keV}$~\cite{Nadler:2019zrb}.  Other
astrophysical constraints from exoplanet~\cite{Leane:2020wob} and
cosmic-ray (CR) down scattering~\cite{Cappiello:2018hsu}, etc.~have
also be considered.

\paragraph{CRDM}
Recently, it was shown that more stringent constraints can be derived
from the elastic scatterings between cosmic-ray (CR) particles and DM
particles.  High-energy CRs in the Galaxy can scatter off halo DM
particles, which results in the energy-loss of
CRs~\cite{Cappiello:2018hsu}, the production of
$\gamma$-rays~\cite{Cyburt:2002uw,Hooper:2018bfw}, and energy-boost of
DM particles~\cite{Bringmann:2018cvk,Ema:2018bih,Cappiello:2019qsw},
etc..  In the last process, a small but irreducible component of DM
(referred to as CRDM) can obtain very high kinetic energies. They can
scatter again off the target nuclei in the detectors of the
underground DM direct detection experiments, and deposit sufficient
energy to cross the detection threshold, which greatly extends the
sensitivity of the current experiments to sub-GeV DM
particles~\cite{Bringmann:2018cvk,Ema:2018bih,Wang:2019jtk,PROSPECT:2021awi,Ge:2020yuf,Cappiello:2019qsw,Dent:2019krz,Bondarenko:2019vrb,Guo:2020oum,Guo:2020drq,Xia:2020apm}.
In this approach the constraints on a constant DM-nucleon
(DM-electron) spin-independent scattering cross section can reach
$\sim10^{-31}(10^{-34})\,\text{cm}^{2}$ for DM particle mass down to
at least $\sim0.1~\text{MeV}$
($\sim1~\text{eV}$)~\cite{Bringmann:2018cvk,Ema:2018bih} (for
constraints on energy-dependent and inelastic scattering cross
sections, see
e.g.~\cite{Dent:2019krz,Bondarenko:2019vrb,Alvey:2019zaa,Guo:2020oum,Su:2020zny,Ema:2020ulo,Bell:2021xff}).
An advantage of this approach is that the obtained constraints are
insensitive to the DM particle mass. This is related to the spectra
feature of the CR flux.  In Ref.~\cite{Xia:2020apm}, it was shown that
if the CR flux follows a power law of $E^{-\alpha}$ with $\alpha=3$,
the DM particle mass dependence in the CRDM flux canceled out
completely, and consequently the derived constraints are DM particle
mass independent. Since the observed CR flux is indeed close to the
$\alpha=3$ case, especially for CR energy above PeV, the constraints
are highly DM particle mass independent and can be extrapolated to
very low DM particles mass far below eV scale. These constraints only
depend on DM-nucleon scattering in the present-day local Universe,
which is complementary to those derived using the observables of the
earlier epochs of the Universe, such as that from
BBN~\cite{Krnjaic:2019dzc},
CMB~\cite{Slatyer:2018aqg,Gluscevic:2017ywp,Boddy:2018kfv,Xu:2018efh},
Lyman-$\alpha$ forest~\cite{Murgia:2018now}, and 21 cm
radiations~\cite{Slatyer:2018aqg}, etc..

For large cross sections above $\mathcal{O}(10^{-30})~\text{cm}^2$,
the DM energy loss due to the same DM elastic scattering off the
nuclei within the crust of the Earth becomes significant. For large
enough cross sections, the DM particle can lose most of its kinetic
energy before reaching the detector, which leads to an exclusion upper
bound (``blind spots'') on the scattering cross section for
underground DM detection experiments. For the leading experiment of
XENON1T with depth around 1.4~km, the typical exclusion upper bound is
around $\mathcal{O}(10^{-28})~\text{cm}^2$ when the effect of the
nuclear form factor is
neglected~\cite{Bringmann:2018cvk,Alvey:2019zaa,Guo:2020drq,Xia:2020apm}. Thus
there exists a gap in the cross section which can be excluded by the
underground DM direct detection experiments and that by the
astrophysical observables. Additional experimental data have to be
considered to close the gap such as using other underground
experiments with shallower sites~\cite{Cappiello:2019qsw}.

\paragraph{This Work}
In this work, we perform a refined analysis on the exclusion bounds on
DM-nucleon scattering cross section $\sigma_{\chi p}$ in this approach
using CR-boosted DM (CRDM).  For the exclusion lower bounds, making
use of the numerical code \galprop{} for CR propagation in the Galaxy,
we for the first time consider the inhomogeneity of the CR
distribution and determine an important parameter of effective
distance $D_\text{eff}$ which is often considered as an free
phenomenological parameter in the literature.  Using spatial-dependent
CR fluxes and including the contributions from the major CR nuclear
species from H to Ni, we obtain $D_\text{eff}\simeq 9~\text{kpc}$ for
CRDM particle with kinetic energy above $\sim 1~\text{GeV}$. Including
CR species heavier than helium leads to a factor of two enhancement of
the total CRDM flux. The corresponding exclusion lower bounds can
reach $\sigma_{\chi p}\sim4\times 10^{-32}~\text{cm}^2$ for DM
particles with mass at MeV scale and below.  For the detection upper
bounds, we numerically simulate the effect of Earth attenuation of
CRDM with the effect of nuclear form factor fully taken into
account. The numerical simulation of Earth attenuation is based on
the \darkprop{} code which is publicly available
at~\cite{DarkProp:v0.1}. We find that for relativistic CRDM particles,
the nuclear form factor significantly reduce the stopping-power of the
Earth's crust. After consistently including the form factors in the
simulation, the exclusion upper bounds are found to be many orders of
magnitude higher than $\mathcal{O}(10^{-28})~\text{cm}^2$ obtained
from the previous analysis~\cite{Bringmann:2018cvk}. Our result closes
the gap between cross sections excluded by the XENON1T experiment and
that by the astrophysical observables such as the CMB, galactic gas
cloud cooling and galactic structure formation.

This paper is organized as follows: In~\Sec{sec:crdm_flux}, we review
the CRDM flux calculation formalism and calculate the effective
distance using spatial-dependent CR flux and including the
contributions from major heavy CR nuclear species.
In~\Sec{sec:earth_attenuation}, we analyze Earth attenuation of CRDM
using both the analytic approach and numerical simulation to obtain
underground CRDM flux.  The predicted direct detection recoil spectra
and corresponding exclusion regions are discussed
in~\Sec{sec:constraints_from_direct_detection_experiment}. We conclude
this work in~\Sec{sec:conclusions}.

\section{CRDM flux}\label{sec:crdm_flux}
\subsection{Kinematics}\label{sub:kinematics}
In this section we briefly review the basic formalism of CRDM flux
calculation. The processes of CR-DM scattering in the galaxy,
DM-nucleus scattering in the crust of the Earth and that within the
underground detector can be approximated as two-body elastic
scattering processes.  In the generic case, for an incident particle
$X$ which elastically scatters off a fixed target particle $Y$, the
recoil kinetic energy $T_Y$ of the particle $Y$ is relate to the
kinetic energy $T_X$ of the particle $X$ as $T_Y = T_Y^{\max}
(T_X)(1-\cos\theta^{*})/2$, where $\theta^*$ is the scattering angle
in the center-of-mass (CM) frame, and the maximal recoil energy of the
target particle $Y$ is given by
\begin{equation}\label{eq:kin_tb_max}
    T_Y^{\max}(T_X)= {\left[1
    + \frac{{(m_Y-m_X)}^2}{2m_Y(T_X+2m_X)}\right]}^{-1}T_X .
\end{equation}
The maximal energy transfer occurs when $m_Y=m_X$, or in the case
where the incident particle X is relativistic with $T_X \gg
m_X^2/2m_Y \ (T_X\gg m_Y/2)$ for $m_X\gg m_Y \, (m_X\ll m_Y)$.  The
minimal energy of the particle $X$ required to produce a recoil energy
$T_Y$ can be obtained by inverting \eq{eq:kin_tb_max}, which gives
\begin{equation}\label{eq:kin_tx_min}
    T_X^{\min} (T_Y)= \left( \frac{T_Y}{2} -
    m_X \right) \left(1 \pm \sqrt{1
    + \frac{2T_Y}{m_Y} \frac{{(m_X+m_Y)}^2}{{(2m_X-T_Y)}^2}}\right) ,
\end{equation}
where the $+(-)$ sign applies for the case of $T_Y > 2m_X \ (T_Y <
2m_X)$.  For instance, let us consider a DM direct detection
experiment with xenon target nuclei ($m_N=122~\text{GeV}$) and a
detection threshold of $T_N^\text{thr} \simeq 5~\text{keV}$.  For
detecting a CRDM particle (denoted as $\chi$) with mass
$m_\chi=1~(10)~\text{MeV}$ in such an experiment, the required minimal
DM kinetic energy, and the corresponding required minimal CR proton
kinetic energy are
\begin{equation}\label{eq:threshold}
	T^\text{thr}_\chi \simeq 16.5~(10.1)~\text{MeV, and } \
	T^\text{thr}_p \simeq 1.92~(0.22)~\text{GeV},
\end{equation}
respectively. Thus for detecting MeV scale DM particles, the required
typical CR proton energy is around the GeV scale.

Neglecting the secondary collisions of CRDM with the interstellar
medium, the CR boosted DM particles should travel in straight lines
before reaching the Earth.  The flux (defined as number of particles
per unite area, time and solid angle, i.e.  $\Phi=dN/dAdtd\Omega$) of
DM particles up-scattered by CR nuclei species $i$ can be written
as~\cite{Bringmann:2018cvk}
\begin{equation}\label{eq:crdm_flux}
    \frac{d \Phi_\chi}{d T_\chi} = \sum_{i} \int_\mathrm{l.o.s}
    d \ell \frac{\rho_\chi(\bm{r})}{m_\chi} \int_{T^{\text{min}}_i} d
    T_i \frac{d \sigma_{\chi i}}{d T_\chi} \frac{d \Phi_i(\bm{r})}{d
    T_i},
\end{equation}
where $\rho_\chi(\bm{r})$ is DM energy density distribution, $m_\chi$
is the mass of DM particle, $d \Phi_i(\bm{r})/d T_i$ is the energy
distribution of the flux of CR species $i$, and $d \sigma_{\chi i}/d
T_\chi$ is the differential cross section for the scattering process.
The integration of DM flux generated from different positions should
be performed along the line-of-sight (l.o.s).

The angular-averaged CRDM flux is defined as $d \bar\Phi_\chi/d
T_\chi \equiv \frac{1}{4\pi}\int d\Omega (d \Phi_\chi/dT_\chi)$, where
$\Omega$ is the solid angle. From \eq{eq:crdm_flux}, the averaged flux
can be written as
\begin{equation}
\label{eq:crdm_flux_k_factor}
\frac{d \bar\Phi_\chi}{d T_\chi}
= \frac{\rho_\chi^\text{loc}}{m_\chi}\sum_{i}
\int d T_i K_i(T_i)
\frac{d \sigma_{\chi i}}{d T_\chi}
\frac{d \Phi_i^\text{loc}}{d T_i} ,
\end{equation}
where the $K_i$ factors which contain all the astrophysical
information are defined as
\begin{equation}
    \label{eq:k_factor}
    K_i(T_i) \equiv \frac{1}{\rho_\chi^{\mathrm{loc}}\frac{d \Phi_i^\mathrm{loc}}{d
    T_i}} \int\frac{d \Omega}{4\pi} \int_\mathrm{l.o.s}
    d \ell \rho_\chi(\bm{r})\frac{d \Phi_i(\bm{r})}{d T_i},
\end{equation}
where the superscript ``loc'' represents the local value in the solar
system.  If the energy spectra of the CR flux do not vary
significantly within the Galaxy, the spatial dependence of the flux
can be factored out as
\begin{align}\label{eq:g-factor}
	\frac{d \Phi_i(\bm{r})}{d T_i} \approx
	g_i(\bm{r})\frac{d \Phi_i^\mathrm{loc}}{d T_i} .
\end{align}
In this case, the $K_i$ factors become nearly energy independent and
only contain the spatial distribution of DM and CR particles.
\begin{equation}\label{eq:kfactor_const}
K_i \approx \int \frac{d\Omega}{4\pi}
\int_\text{l.o.s} \frac{\rho_{\chi}(\bm{r})}{\rho_\chi^\text{loc}}
g_i(\bm{r}) d \ell .
\end{equation}
Alternatively, the CRDM flux can be rewritten as
\begin{equation}
    \label{eq:crdm_flux_d_Tchi} \frac{d \bar\Phi_\chi}{d T_\chi} =
    D_\text{eff}(T_\chi) \frac{\rho_\chi^\text{loc}}{m_\chi} \sum_{i} \int
    dT_i \frac{d \sigma_{\chi i}}{d
    T_\chi} \frac{d \Phi_i^\text{loc}}{d T_i},
\end{equation}
where the quantity $D_\text{eff}(T_\chi)$ is defined as
\begin{equation}
    \label{eq:deff} D_\text{eff}(T_\chi) \equiv \frac{\sum_{i}\int d
    T_i K_i(T_i)\frac{d \sigma_{\chi i}} {d
    T_\chi}\frac{d \Phi_i^\text{loc}}{d T_i}} {\sum_{i}\int d
    T_i\frac{d\sigma_{\chi i}}{d T_\chi} \frac{d \Phi_i^\text{loc}}{d
    T_i}}
\end{equation}
which in general depends on the kinetic energy $T_\chi$. In the
simplified assumption where the CR flux in the Galaxy is almost the
same as the local CR flux, namely, $g_i(\mathbf{r})\approx 1$, all the
$K_i$ factors become the same, i.e., $K_i\approx K$.  In this case,
the effective distance $D_\text{eff}(T_\chi)$ becomes energy
independent as well, and reduces to a constant
$D_\text{eff}(T_\chi)\approx D_\text{eff}$ which is given by
\begin{equation}\label{eq:deff_const}
    D_\text{eff} \approx K
    = \int \frac{d\Omega}{4\pi} \int_\text{l.o.s} \frac{\rho_{\chi}(\bm{r})}{\rho_\chi^\text{loc}}
    d \ell,
\end{equation}
which is the so-called effective distance and commonly adopted in the
literature~\cite{Bringmann:2018cvk}. $D_{\text{eff}}
= \mathcal{O}(1-10)~\text{kpc}$ is often chosen as a benchmark
value. The deviation of effective distance $D_\text{eff}(T_\chi)$ from
a constant directly reflects the effect of the inhomogeneity of the CR
flux in CRDM flux.

\subsection{Calculation of the effective distance}\label{sub:calculation_of_the_effective_distance}

In most of the recent analyses on CRDM, the value of $D_\text{eff}$
was taken as a parameter roughly in the range of $1-10~\text{kpc}$,
which becomes a source of uncertainty in phenomenological
analysis. This quantity is actually calculable as the CR distribution
within the Galaxy is well modeled.  In this work, we focus on the CRDM
particles with mass around $\mathcal{O}(\text{MeV}-\text{GeV})$. This
sub-GeV region recently received significant amount of attention as
they can be potentially reached by the current DM direct detection
experiments with lower thresholds and improved analysis methods.  For
MeV-GeV scale CRDM particles which are relevant to the DM direct
detection, the dominant contribution to their flux comes from CRs with
energy below PeV which are mainly of Galactic origin.  Our Galaxy is
filled with random magnetic fields with typical strength around a few
$\mu G$. Charged CR nuclei produced within the Galaxy are trapped by
the magnetic fields and in diffusive random motion for a long period
of $\mathcal{O}(\text{Myrs})$ before escaping from the Galaxy. The
distribution of Galactic CRs are well described by diffusion models
with parameters determined by local CR measurements.

The propagation of CR particles through the Galaxy can be approximated
by a diffusion model. The diffusion halo is parameterized by a
cylinder with radius $R\simeq20~\text{kpc}$ and half-height
$Z_\mathrm{h}=1\sim10~\text{kpc}$.  The diffusion equation for the CR
charged particles reads~\citep{Berezinsky:1990qxi,Strong:2007nh}
\begin{equation}
	\frac{\partial\psi}{\partial
	t}=\nabla\cdot(D_{xx}\nabla\psi-\boldsymbol{V}_{c}\psi)+\frac{\partial}{\partial
	p}p^{2}D_{pp}\frac{\partial}{\partial
	p}\frac{1}{p^{2}}\psi-\frac{\partial}{\partial
	p}\left[\frac{dp}{dt}\psi-\frac{p}{3}(\nabla\cdot\boldsymbol{V}_{c})\psi\right]-\frac{1}{\tau_{f}}\psi-\frac{1}{\tau_{r}}\psi+q(\boldsymbol{r},p),\label{eq:propagation}
\end{equation}
where $\psi(\boldsymbol{r},p,t)$ is the CR number density per unit
momentum, $D_{xx}$ is the spatial diffusion coefficient, and
$\boldsymbol{V}_{c}$ is the convection velocity. The re-acceleration
effect is described as diffusion in momentum space and is determined
by the coefficient $D_{pp}$. The quantity $dp/dt$ stands for the
momentum loss rate. $\tau_{f}$ and $\tau_{r}$ are the time scales for
fragmentation and radioactive decay respectively, and
$q(\boldsymbol{r},p)$ is the source term.  The convection velocity
$\boldsymbol{V}_{c}$ is assumed linearly increases with the distance
$z$ from the galactic plane with the gradient $dV_c/dz$.  The
energy-dependent spatial diffusion coefficient $D_{xx}$ is
parameterized as $D_{xx}=\beta^\eta
D_{0}{\left({\rho}/{\rho_{0}}\right)}^{\delta}$, where $\rho=p/Ze$ is
the rigidity of CR particles with electric charge $Ze$, $\delta$ is
the spectral power index which can have different values
$\delta=\delta_{1(2)}$ for $\rho$ below (above) a reference rigidity
$\rho_{0}$ (4~GV in this work). $D_{0}$ is a normalization constant,
$\beta=v/c$ is the velocity of CR particles and $\eta\approx 0.7$.
The momentum diffusion coefficient $D_{pp}$ is related to $D_{xx}$ as
$D_{pp}D_{xx}=4V_{a}^{2}p^{2}/(3\delta(4-\delta^{2})(4-\delta))$,
where $V_{a}$ is the Alfv\'en velocity of disturbances in the
hydrodynamical plasma~\citep{Berezinsky:1990qxi}.  The spatial
boundary conditions are set by assuming that free particles escape
beyond the halo, i.e., $\psi(R,z,p)=\psi(\boldsymbol{r},\pm
Z_\mathrm{h},p)=0$. The steady-state solution can be obtained by
setting $\partial\psi/\partial t=0$.

We adopt the numerical
code \galpropv{}~\cite{Strong:1998pw,Moskalenko:1997gh} to solve the
CR diffusion equation with given source distributions and boundary
conditions. The \galprop{} code uses real Milky Way information
including the gas distribution, magnetic field distribution, etc., and
can calculate the CR species from proton to nickel with related
isotopes included.  Recently, the \helmod{}
code~\cite{Boschini:2017fxq} has been developed to describe the solar
modulation and derive the low energy local interstellar (LIS) flux of
CR $\bar{p}$, $e^{-}$, and nuclei $Z \le 28$~\cite{Boschini:2020jty}.
The \helmod{} group has performed a fit to the CR nuclei $_1$H $-$
$_{28}$Ni with CR data up to $200~\text{TeV}$ from
Voyager~1~\cite{Cummings:2016pdr}, ACE-CRIS~\cite{ACE:2001},
CREAM~\cite{Ahn:2008my,Ahn:2009tb},
NUCLEON~\cite{Gorbunov:2018stf,Grebenyuk:2018jcb},
CALET~\cite{CALET:2019bmh}, HEAO-3-C2~\cite{Engelmann:1990zz},
DAMPE~\cite{DAMPE:2019gys}, and AMS-02~\cite{AMS:2018tbl}.  The
best-fit values of the main propagation parameters are summarized
in~\tab{tab:propagation_parameter}.
\begin{table}[t]
    \centering \begin{tabular}{ccccc} \toprule $Z_\mathrm{h}$ [kpc] &
    $D_0$ [$10^{28}$ cm$^2$ s$^{-1}$] & $\delta$ & $V_a$ [km s$^{-1}$]
    & $d V_\text{c}/d z$ [km s$^{-1}$ kpc$^{-1}$] \\ \colrule 4.0 &
    4.3 & 0.415 & 30 & 9.8 \\ \botrule \end{tabular} \caption{Main
    propagation parameters of the CR diffusion equation from
    Ref.~\cite{Boschini:2020jty}.}\label{tab:propagation_parameter}
\end{table}

The primary source term $q(\bm{r}, p)$ can be factorized into the
product of a spatial and rigidity distribution function, namely,
$q_\text{prim}(r, z, p) \propto f(r, z) q(R)$. The energy-dependent
part takes the form of a smoothly broken power law in rigidity as
follows
\begin{equation}\label{eq:inj}
    q(R) \propto
    {\left(\frac{R}{R_0}\right)}^{-\gamma_0} \prod_{i=0}^2 {\left[
    1+{\left(\frac{R}{R_i}\right)}^{\frac{\gamma_i
    - \gamma_{i+1}}{s_i}} \right]}^{s_i} ,
\end{equation}
where $s_i$ is negative (positive) for $|\gamma_i| > |\gamma_{i+1}|$
($|\gamma_i| < |\gamma_{i+1}|$).  The spectral parameters $R_i$,
$\gamma_i$ and $s_i$ for each CR species are tuned to reproduce the
experiment data.  Recently, the \helmod{} group updated the source
spectral parameters~\cite{Boschini:2021gdo,Boschini:2021ekl} to
reproduce the new AMS-02 data of CR iron~\cite{AMS:2021lxc} and
fluorine~\cite{AMS:2021tnd}. The primary source spectral parameters
are summarised in
Appendix~\ref{sec:parameters_of_the_primary_cr_sources}.  The spatial
distribution of primary source $f(r, z)$ is assumed to follow the
standard pulsar distribution~\cite{Boschini:2020jty,Lorimer:2006qs}
\begin{equation} \label{eq:primary_source_spatial}
    f(r, z) \propto
    {\left(\frac{r}{r_{\odot}}\right)}^\alpha \exp \left(-\beta \frac{r-r_{\odot}}{r_{\odot}}
    - \frac{|z|}{z_0}\right),
\end{equation}
where $r_{\odot} = 8.5~\text{kpc}$, $\alpha = 1.9$, $\beta = 5.0$, and
$z_0 = 0.18~\text{kpc}$.  The primary source term for each CR species
is normalized to the proton source according to their abundances which
are provided in Ref.~\cite{Boschini:2020jty}. An overall normalization
is introduced after the calculation of CR propagation to fit the
experimental data. We take these two normalization energy at 100
GeV/nucleon with the proton flux at the normalization energy set to
$4.47\times
10^{-9}~\text{MeV}^{-1}\text{cm}^{-2}\text{s}^{-1}\text{sr}^{-1}$.

\begin{figure}[tbp]
	\centering \includegraphics[width=0.49\linewidth]{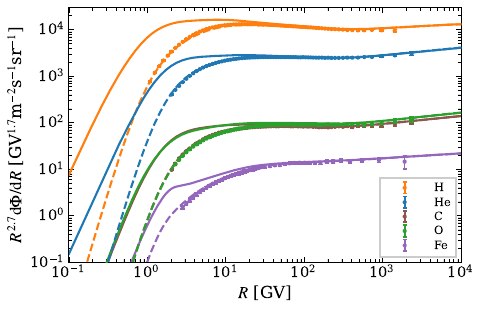} \includegraphics[width=0.49\linewidth]{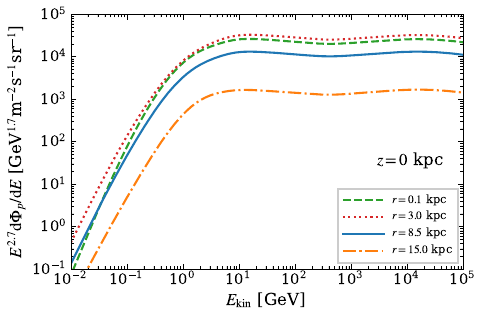} \caption{\label{fig:cr_flux_ams}
	(Left) CR flux as a function of rigidity for nuclear species
	of H, He, C, O, and Fe, together with the AMS-02
	data~\cite{AMS:2015tnn,AMS:2015azc,AMS:2017seo,AMS:2021lxc}.
	The solid lines represent LIS fluxes of CRs calculated
	using \galpropv{}, and the dashed lines indicate the
	corresponding solar modulated fluxes calculated
	using \helmodv{}.  (Right) CR proton flux as a function of
	kinetic energy in the galactic plane (z = 0) at different
	distances to the Galactic Center $r = 0.1$, $3.0$, $8.5$, and
	$15.0~\text{kpc}$.}\end{figure}

For the DM energy distribution, we adopt the standard NFW
profile~\cite{Navarro:1995iw} which is parametrized
as~\cite{Ackermann:2012rg}
\begin{equation}
    \rho^\mathrm{NFW}_\chi(r) = \rho_\chi^\text{loc} {\left(1
    + \frac{R_{\odot}}{R_\text{s}} \right)}^2
    {\left(\frac{r}{R_{\odot}}\right)}^{-1} {\left(1
    + \frac{r}{R_\text{s}} \right)}^{-2}, \label{eq:nfw}
\end{equation}
where $R_\text{s} = 20~\text{kpc}$, $R_{\odot} = 8.5~\text{kpc}$, and
$\rho_\chi^\text{loc} = 0.3~\text{GeV cm}^{-3}$.  The variations of
CRDM flux derived from different DM density profiles are typically at
a few percent level, and choosing the NFW profile leads to more
conservative CRDM flux. A detailed comparison of the effective
distance of CRDM in different DM density profiles can be found in
Appendix~\ref{sec:comparison_of_dark_matter_profiles}.

\paragraph{Figure 1}
In the left panel of \fig{fig:cr_flux_ams}, we show the local
interstellar (LIS) fluxes of the CR nuclear species of H, He, C, O,
and Fe which contribute up to $\sim 78\%$ of the total CRDM flux for a
typical DM particle with $m_\chi=10~\text{MeV}$. The LIS fluxes are
calculated using \galpropv{}. The effect of solar modulation
calculated using \helmodv{} are also shown. The results show a good
agreement with the local CR measurement such as the recent AMS-02
data.  In the right panel of \fig{fig:cr_flux_ams}, we show the CR
proton fluxes in the galactic plane ($z=0$) at difference distances to
the Galactic Center. It can be seen that all the CR fluxes at
different distances have similar spectral shapes, the major
differences are only in the overall normalization, which suggests
that \eq{eq:g-factor} is a reasonable approximation for the CR
distribution within the Galaxy.

\begin{figure}[tbp]
	\centering \includegraphics[width=0.49\linewidth]{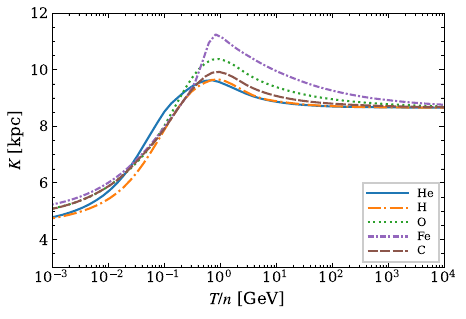} \includegraphics[width=0.49\linewidth]{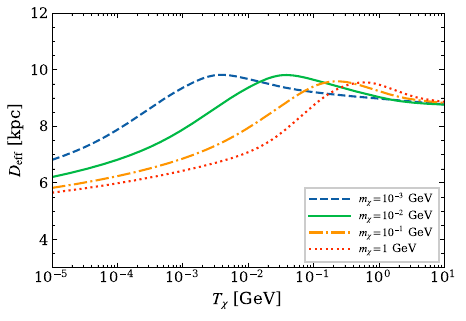} \caption{(Left)
	$K$ factors as a function of kinetic energy per nucleon for CR
	nuclei H, He, C, O, and Fe.  (Right) Effective distance
	$D_\text{eff}$ as a function of DM kinetic energy with
	different DM particles masses from $1~\text{MeV}$ to
	$1~\text{GeV}$.}\label{fig:d_factor}
\end{figure}

\paragraph{Figure 2}
In the left panel of~\fig{fig:d_factor}, the values of the $K$ factors
calculated from~\eq{eq:k_factor} for the same CR nuclei as that
in~\fig{fig:cr_flux_ams} are shown. The energy dependencies of the $K$
factors for different CR species are quite similar: starting from
about $4.5~\text{kpc}$ at low energy below
$10^{-3}~\text{GeV}/\text{nucleon}$, reaching a peak around
$10~\text{kpc}$ when the kinetic energy is $\sim
1~\text{GeV}/\text{nucleon}$, and then decreasing to a constant about
$8.7~\text{kpc}$ at high energy above $\sim
10~\text{GeV}/\text{nucleon}$. In the energy range of
$10^{-3}-10^{3}~\text{GeV}$, the overall change in the $K$ factors is
about a factor of two.  The asymptotic behavior is related to the fact
that at high (low) energy region with kinetic energy above
$10~\text{GeV}/\text{nucleon}$ (below 1~GeV/nucleon) the local and LIS
CR spectrum follow the same single power law, thus the energy
dependence almost cancels out.  Unlike the $K$ factors which are
defined purely by astrophysical quantities, the effective distance
$D_\text{eff}$ essentially depends on the DM-nucleus scattering cross
section and DM particle mass. Since the effective distance is an
average of $K$ factors weighted by the relevant cross sections and
local CR fluxes for a given $T_\chi$, the variation in $D_\text{eff}$
will not exceed that in the value of the $K$ factors.  For large
$T_\chi \gtrsim~\text{GeV}$, the value of $D_\text{eff}$ approaches a
constant of $D_\text{eff}\simeq 9~\text{kpc}$, which is very weakly
dependent of $m_\chi$.

The CR nucleus-DM scattering cross section $\sigma_{\chi i}$ can be
related to that at nucleon level $\sigma_{\chi p}$. In this work we
consider spin-independent scattering and adopt the simplified
assumption that the DM-nucleon scattering is isospin-conserving. In
this case the DM-nucleus scattering cross section at zero momentum
transfer $\sigma^0_{\chi i}$ is related to DM-nucleon scattering cross
section $\sigma_{\chi p}$ which is to be constrained by DM direct
detection experiments as~\cite{Jungman:1995df}
\begin{equation}\label{eq:sigma_chii_chip}
	\sigma_{\chi i}^0 = A_i^2 \frac{\mu_{\chi i}^2}{\mu_{\chi
	p}^2} \sigma_{\chi p},
\end{equation}
where $A_i$ is the atomic mass number of the nucleus $i$, and
$\mu_{ab}=m_a m_b/(m_a+m_b)$ is the two-particle reduced mass, The CR
nucleus-DM scattering cross section is given by
\begin{equation}\label{eq:sigma_chi_nuclei}
	\frac{d\sigma_{\chi i}}{dT_\chi}=\frac{F_i^2(q^2) A_i^2
	}{T_\chi^{\max}(T_i)}\frac{\mu_{\chi i}^2}{\mu_{\chi
	p}^2} \sigma_{\chi p},
\end{equation}
where $F_i(q^2)$ is the nuclear form factor at the momentum transfer
$q^2 = 2m_\chi T_\chi$.  We emphasize that the total cross section
$\sigma_{\chi i}$ from integrating \eq{eq:sigma_chi_nuclei} is in
general \textit{different} from $\sigma^0_{\chi i}$ after the effect
of form factor is taken into account.

\subsection{CRDM flux}\label{sub:crdm_flux}
Using the expression of $T^\text{min}_i$ in \eq{eq:kin_tx_min} and the
cross section in \eq{eq:sigma_chi_nuclei}, the angular averaged flux
of CRDM can be calculated as the sum of the contributions from the
individual CR species
\begin{equation}\label{eq:flux_crdm_elastic}
    \frac{d \bar\Phi_\chi}{d T_\chi}
    = \frac{\rho_\chi^\text{loc}}{m_\chi} \sum_i F_i^2(2m_\chi
    T_\chi) \sigma_{\chi i}^{0} \int_{T_i^{\min}(T_\chi)}^{\infty} d
    T_i \frac{K_i(T_i)}{T_\chi^{\max}} \frac{d \Phi_i^\text{loc}}{d
    T_i}.
\end{equation}

For proton and helium, we take the dipole form factor $ F_i(q^2) = 1 /
{(1 + q^2/\Lambda_i^2)}^2$~\cite{Perdrisat:2006hj}, where
$\Lambda_p \approx 770~\text{MeV}$ and $\Lambda_{\text{He}} \approx
410~\text{MeV}$.  For heavier nuclei we adopt the conventional Helm
form factor~\cite{Helm:1956zz,Lewin:1995rx}
\begin{equation}\label{eq:helmformfactor}
	F(q^2) = \frac{3j_1(qR_1)}{qR_1} e^{- \frac{1}{2} q^2 s^2} ,
\end{equation}
where $j_1(x)$ is the spherical Bessel function of the first kind,
$R_1 = \sqrt{R_A^2-5s^2}$ with $R_A \approx 1.2 A^{1/3}~\text{fm}$,
and $s\approx 1~\text{fm}$.

\begin{figure}[tbp]
	\centering \includegraphics[width=0.49\textwidth]{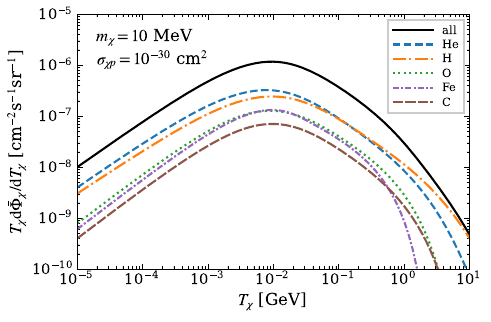}\includegraphics[width=0.49\textwidth]{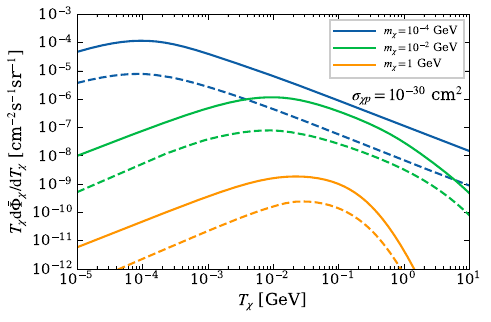}\caption{(Left)
	CRDM fluxes derived from the five most dominant CR species of
	He, H, O, Fe, and C with DM particle mass $m_\chi =
	10~\text{MeV}$ and cross section $\sigma_{\chi p} =
	10^{-30}~\text{cm}^2$.  The black solid line indicates the
	total flux contributed by all the CR species with $Z \leq 28$.
	(Right) Total CRDM fluxes contributed by all the CR species
	for different DM particle masses $m_\chi = 10^{-4}$,
	$10^{-2}$, and $1~\text{GeV}$ (solid lines) with cross section
	$\sigma_{\chi p} = 10^{-30}~\text{cm}^2$.  The CRDM fluxes
	derived from the parameterised LIS fluxes of proton and helium
	in~\cite{Boschini:2017fxq} with a constant effective distance
	$D_\text{eff} = 1~\text{kpc}$ in Ref.~\cite{Bringmann:2018cvk}
	are also shown for comparison (dashed
	lines).}\label{fig:crdm_flux}
\end{figure}

\paragraph{Figure 3}
In the left panel of \fig{fig:crdm_flux}, the fluxes of CRDM
contributed by the five most dominant CR species are shown and
compared with the total flux contributed by all the CR species, for a
typical light CRDM particle with $m_\chi=10~\text{MeV}$. Among the CR
species, helium contribution is the most dominant, which contributes
about $28\%$ of the total CRDM flux. Helium and proton together
contribute about $50\%$ of the total flux.  Despite much smaller
fluxes, contributions from heavier CR nuclei can be comparable to that
from helium/proton due to the enhancement factor $A_i^2$ in the
spin-independent scattering cross section. Including the contributions
from oxygen, iron and carbon etc.~leads a factor of two enhancement in
$D_\text{eff}$.  The detailed contribution from each CR species are
summarized in~\tab{tab:percentage}.  Note, however, that for heavier
DM particles, the total contribution is still dominated by helium and
proton due to the stronger form factor suppression. For instance, for
$1~\text{GeV}$ DM particle, the CR proton contributes $47\%$ of the
CRDM flux and helium contributes $39\%$.  During the calculation, we
neglect the light CRs lithium and beryllium as they are secondary CRs
with very low abundances.
\begin{table}[tb]
	\begin{tabular}{lcc} \toprule nucleus & CRDM flux
		[cm$^{-2}$s$^{-1}$] & percentage [\%] \\ \colrule He &
		$1.77\times 10^{-5}$ & 28.6 \\ H & $1.38\times
		10^{-5}$ & 22.3 \\ O & $6.61\times 10^{-6}$ & 10.7 \\
		Fe & $6.19\times 10^{-6}$ & 10.0 \\ C & $3.60\times
		10^{-6}$ & 5.84 \\ Si & $2.76\times 10^{-6}$ & 4.48 \\
		Mg & $2.68\times 10^{-6}$ & 4.35 \\ Ne & $1.24\times
		10^{-6}$ & 2.01 \\ N & $1.07\times 10^{-6}$ & 1.74 \\
		sum & $5.57\times 10^{-5}$ &
		90.0 \\ \botrule \end{tabular} \caption{Contributions
		to the total CRDM flux from different CR species.  The
		DM particle mass is fixed at $m_\chi = 10~\text{MeV}$,
		and the flux is integrated over the DM kinetic energy
		range $10^{-5} - 10~\text{GeV}$ and $4\pi$ solid
		angle.  The species which contribute up to $90\%$ of
		the total CRDM flux are listed.}\label{tab:percentage}
\end{table}

\paragraph{Figure 3}
The total CRDM fluxes for different DM particle masses are shown in
the right panel of \fig{fig:crdm_flux}. We compare the CRDM flux of
different DM particle masses with that derived using a constant
$D_\text{eff} = 1~\text{kpc}$~\cite{Bringmann:2018cvk} in the right
panel of~\fig{fig:crdm_flux}.  For lighter DM, the flux exhibits a
power-law behavior in the region of interest, which is related to the
approximate power-law CR spectrum and constant cross section. For
heavier DM, the nuclear form factor leads to a steeper decrease in the
high-energy part of the flux.  The shape of the spectra does not
change much after the inhomogeneous CR fluxes are taken into account
as $D_\text{eff}(T_\chi)$ changes less than a factor of two for the
energy range shown in the right panel of~\fig{fig:d_factor}.  The
slope change that can be seen below $10^{-2}~\text{GeV}$ for $m_\chi =
1~\text{GeV}$ is due to the fact that the parameterization of the LIS
CR fluxes adopted from~\cite{Boschini:2017fxq} are only valid for CRs
with rigidity higher than $0.2~\text{GV}$, and the CRDM fluxes
contributed by lower energy CRs are ignored. The kinetic energy of the
LIS CRs in this work are calculated down to 1 eV\note{?} per nucleon
by~\galpropv{}.  We find that the CRDM fluxes are enhanced by more
than an order of magnitude compared with the benchmark model with
$D_\text{eff}=1~\text{kpc}$~\cite{Bringmann:2018cvk}.  In the whole
enhancement, roughly a factor of $\sim8$ enhancement arises from the
value of $D_\text{eff}$ and a factor of $\sim2$ from CR species
heavier than proton and helium.

Although in this work we are only interested in the angular-averaged
CRDM flux, the CRDM flux at the surface of the Earth is highly
anisotropic in the arrival direction. CRDM flux from the direction of
the galactic center (GC) is significantly higher than that from the
anti-GC (AC) direction. For instance, for DM particles with mass
$m_\chi = 10~\text{MeV}$ and scattering cross section $\sigma_{\chi p}
= 10^{-30}~\text{cm}^2$, the CRDM flux with $T_\chi \geq 0.01$~MeV and
integrated within a $5^\circ$ cone around the directions of GC and AC
are $\Phi_\text{GC} = 3.1\times10^{-6}~\text{cm}^{-2}\text{s}^{-1}$
and $\Phi_\text{AC} = 2.6\times10^{-8}~\text{cm}^{-2}\text{s}^{-1}$,
respectively.  The final anisotropy measured at the underground
detectors will be significantly smaller due to the multiscatterings
between the DM particles and nucleus within the Earth. The remaining
anisotropy can still leads to non-negligible diurnal modulation of the
event rate, which can be used to set stringent constraints on the
property of CRDM particles~\cite{Ge:2020yuf,PandaX-II:2021kai}.

\section{Earth Attenuation}\label{sec:earth_attenuation}
Before arriving at the underground detectors, CRDM particles lose
energy due to the similar elastic scattering off the nuclei in the
interstellar gas, the atmosphere and the crust of the Earth. For the
cross sections under consideration, the dominant energy-loss process
is the scattering off the nuclei within the Earth crust.  The energy
loss can be estimated using the analytic approach which is based on
the one-dimensional collision
approximation~\citep{Starkman:1990nj,Kouvaris:2014lpa}. A more
sophisticated approach is to use Monte-Carlo simulations to directly
simulate the three-dimensional motion and multiple collisions of DM
particles inside the Earth.  In this work, both approaches are adopted
and compared. We focus on the impact of the form factor on the energy
spectrum of CRDM deep underground. For simplicity, we only consider
elastic scattering which is an irreducible process.

\subsection{Analytic approach}\label{sub:analytic_approach}
In the approximate analytic approach, the decrease of $T_{\chi}$ with
respect to depth $z$ due to the elastic scattering off the nucleus $N$
in Earth's crust is given by
\begin{equation}\label{eq:attenuation}
	\frac{d T_\chi}{d z} = - \sum_N
	n_N \int_0^{T_N^{\max}} \frac{d \sigma_{\chi N}}{d T_N} T_N d
	T_N,
\end{equation}
where $n_N$ is the number density of nucleus species $N$. The maximal
recoil energy can be obtained from \eq{eq:kin_tb_max}, and the
differential cross section from \eq{eq:sigma_chi_nuclei} using the
appropriate substitutions.  Since the dominant chemical component of
the crust is oxygen with $m_N\approx 15~\text{GeV}$, we focus on the
case where $m_\chi, T_\chi \ll m_N$.  Using the expression of
$T^{\text{max}}$ in \eq{eq:kin_tb_max}, the energy-loss with respect
to the distance to the surface $z$ can be rewritten as
\begin{align}\label{eq:dTdz}
	\frac{d T_\chi}{d
	z}=-\frac{1}{\ell_E(T_\chi)}\left(T_\chi+ \frac{T_\chi^2}{2
	m_\chi}\right) ,
\end{align}
where $\ell_{E}$ is the mean free path for energy loss, and
$\ell_{E}^{-1}=2 m_\chi\sum_N g_N(T_\chi) n_N \sigma^0_{\chi N}/m_N$.
The factor $g_N$ in the expression of $\ell_E$ which contains the
effect of the form factor is given by
\begin{equation}
    g_N(T_\chi)=\int^{T^{\text{max}}_N}_{0}
    F^2_N(q^2)\frac{2T_N}{{(T^{\text{max}}_N)}^2} dT_N.
\end{equation}
Note that in the generic case the mean free path depends on $T_\chi$
through the factor $g_N(T_\chi)$.

We start with an oversimplified case where $F_N(q^2)\to 1$. In this
case the factor $g_N$ approaches unity, as expected.  In this simple
case the \eq{eq:dTdz} can be integrated out analytically. The solution
for $T_\chi(z)$ at depth $z$ is related to that at surface $T_\chi(0)$
as follows
\begin{equation}\label{eq:Tz_noformfac}
	T_\chi(z)=\frac{T_\chi(0)
	e^{-z/\ell_E^0}}{1+\frac{T_\chi(0)}{2
	m_\chi}\left(1-e^{-z/\ell_E^0}\right)} ,
\end{equation}
where $\ell_E^0$ is the mean free path for energy loss in the case of
$F_N(q^2)=1$. For isospin-conserving scattering,
${(\ell_E^0)}^{-1}\approx 2 m_\chi \rho_\oplus \sigma_{\chi p}/m_p^2$
where $\rho_\oplus=2.7~\text{g}\cdot\text{cm}^{-3}$ is the averaged
mass density of the outer crust of the Earth, and $m_p$ is the proton
mass. Note that in this special case $\ell_E^0$ is independent of the
chemical composition of the crust~\cite{Xia:2020apm}.

For relativistic CRDM particles, the energy loss term in \eq{eq:dTdz}
is dominantly proportional to $T^2_{\chi}$. Thus for higher energy DM
particles, the energy loss is more efficient.  Due to the fast energy
loss, if the CRDM particles remain to be relativistic at depth $z$
with $z\ll \ell^0_E$, the CRDM flux will be cutoff at a maximal energy
as suggested by \eq{eq:Tz_noformfac}. For isospin-conserving
scatterings, the cutoff energy can be approximated as
\begin{equation}\label{eq:Tz_relatv}
T^\text{cut}_\chi(z) \approx \frac{m_p^2}{z \rho_\oplus \sigma_p} .
\end{equation}
Note that it is independent of the initial value of $T_\chi(0)$ at the
surface and the DM particle mass.  For a given experiment with a
detection threshold $T_N^\text{thr}$, there exists a threshold in the
DM energy $T_\chi^\text{thr}$ below which the DM particle cannot
register a recoil signal. If $T^\text{cut}_\chi(z) $ is below this
threshold, then the DM particles become undetectable.  From this
cutoff, one can set a upper bound on the cross section $\sigma_{\chi
p}$ that can be reached by the experiment under consideration.  Using
the benchmark values given in \eq{eq:threshold}, for
$z=1.4~\text{km}$, the upper bounds is $\sigma_{\chi p}\sim 2.5\times
10^{-28}~\text{cm}^2$ which is highly insensitive to the DM particle
mass.  This result is in an excellent agreement with the findings in
the previous analysis in~\cite{Bringmann:2018cvk}.  If the CRDM
particles become non-relativistic at depth $z$. The cutoff will mildly
depend on the DM initial kinetic energy $T_\chi(0)$ at the surface,
and the DM energy decreases with depth exponentially, i.e.,
$T_\chi(z)\propto T_\chi(0)e^{-z/\ell_E}$.

\paragraph{Formfactor}
In the next step, we turn on the effect of nuclear form factors. For
CRDM particles with energy above $\sim$ GeV, the typical momentum
transfer $q$ in the scattering off the nuclei within the crust can
easily be above a few tens of MeV, which suggests that for high energy
CRDM, the total cross section $\sigma_{\chi N}$ can be significantly
smaller than $\sigma_{\chi N}^0$.  The ratio $R$ between the two cross
sections is given by
\begin{equation}
    \label{eq:sigma_ratio} R(T_\chi)\equiv\frac{\sigma_{\chi
	N}}{\sigma^0_{\chi N}}
	=\int_0^{T_N^\text{max}(T_\chi)} \frac{F^2(q^2)}{T_N^\text{max}(T_\chi)}
	dT_N ,
\end{equation}
where $q^2=2 m_N T_N$. For typical values of $T_\chi=100~\text{MeV}$,
1~GeV and 10~GeV, the corresponding ratios with $m_\chi =
10~\text{MeV}$ and $m_N \approx 15~\text{GeV}$ for oxygen are
$R=0.40$, $5.7\times 10^{-3}$ and $1.2\times10^{-4}$,
respectively. Thus for CRDM with energy above GeV, the total cross
section is suppressed by more than two orders of magnitude.
Consequently, the energy loss due to the Earth attenuation can be much
smaller.

The Helm form factor in \eq{eq:helmformfactor} is in general an
oscillating function. However, before reaching the first zero of the
Bessel function $\zeta_1=4.449$, i.e, $q R_1\leq \zeta_1$, the Helm
form factor can be well approximated as a Gaussian form factor
\begin{equation}
    {F(q^2)}_N\approx e^{-q^2/\Lambda_N^2} \quad (\text{for }q R_1
    < \zeta_1),
\end{equation}
where $\Lambda_N^{-2} \approx R_1^2/a^2+s^2/2$ and $a\approx
3.2$\note{?}.  For oxygen nuclei $\Lambda_N\approx
0.207~\text{GeV}$. In this case, the evolution equation can be
approximated as
\begin{equation}
		\frac{d T_\chi}{d z} \approx -\frac{1}{2}\sum_N
		n_N \sigma_{\chi N}^0 T^{\text{max}}_N f(4 m_N
		T_N^\text{max}/\Lambda^2_N) ,
\end{equation}
where the function $f(x)=2[1-(1+x)e^{-x}]/x^2$ describes the effect of
the form factor. It is evident that $f(x)\to 1$ for $x\to 0$.  For
large $x$, the function has the asymptotic behavior of $f(x)\to
2/x^2$.  Thus the solution for large $x$ is
\begin{equation}\label{eq:Tz_highenergy}
	T^3_\chi(z) \approx T^3_\chi(0)-\frac{3z}{32}\sum_N
	n_N \sigma_{\chi N}^0 \frac{\Lambda^4_N}{m_N} .
\end{equation}
The second term in the r.h.s of the above equation is a relatively
small quantity. It is less than ${(0.15~\text{GeV})}^3$ for
$\sigma_{\chi p}\leq 10^{-28}~\text{cm}^2$ at $z=1.4~\text{km}$. Thus
in a good approximation $T_\chi(z)\approx T_\chi(0)$ for CRDM
particles with energy above GeV, which suggests that energy loss is
negligible for high energy CRDM particles.

\begin{figure}[tbp]
\includegraphics[width=0.7\linewidth]{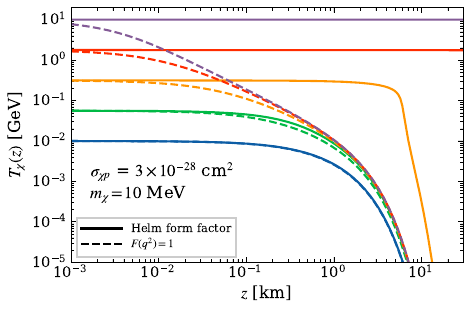}
	\caption{Solution of the energy loss equation
        of~\eq{eq:attenuation} as a function of depth $z$ for
        different initial DM kinetic energies at the surface of the
        Earth ($z = 0$).  The DM particle mass and the cross section
        are set to be $m_\chi = 10~\text{MeV}$ and $\sigma_{\chi p} =
        3\times10^{-28}~\text{cm}^{2}$, respectively.  The results
        calculated with the effect of the nuclear form factor
        considered (neglected) are shown with solid (dashed)
        lines.}\label{fig:t_z_ballistic}
\end{figure}

\paragraph{Figure 4}
The values of the underground CRDM kinetic energy $T_\chi(z)$ as a
function of depth $z$ from numerically solving the \eq{eq:attenuation}
with the Helm form factor included are shown
in \fig{fig:t_z_ballistic} for different initial values of $T_\chi(0)$
at surface.  For simplicity, we use the uniform density Earth model as
in Ref.~\cite{Emken:2018run}, with the relative chemical element
abundances listed in~\tab{tab:earth_crust}.  In the figure we consider
a typical DM particle mass $m_\chi=10~\text{MeV}$ and cross section
$\sigma_{\chi p}=3\times 10^{-28}~\text{cm}^2$. The analytic solutions
without including the form factor from \eq{eq:Tz_noformfac} are also
shown for comparison.  For small initial value $T_\chi(0) \lesssim
0.1~\text{GeV}$, the two approaches agree with each other quite
well. The CRDM energy drops rapidly after $z \gtrsim
1~\text{km}$. However, for higher energy CRDM particles, the effect of
the form factor becomes significant. Including the form factor
essentially reduces the total cross section, which result in smaller
energy loss. For CRDM above GeV, the energy loss is so small that the
DM particle energy is almost unchanged at depth $z\sim$~km. The
numerical results in \fig{fig:t_z_ballistic} confirm the approximate
analytic result in \eq{eq:Tz_highenergy}.
\begin{table}[bp]
	\centering \begin{tabular}{ccccccccc} \toprule Element &
	$^{16}$O & $^{28}$Si & $^{27}$Al & $^{56}$Fe & $^{40}$Ca &
	$^{39}$K & $^{23}$Na & $^{24}$Mg \\ \colrule [wt\%] & 46.6 &
	27.7 & 8.1 & 5.0 & 3.6 & 2.8 & 2.6 &
	2.1 \\ \botrule \end{tabular} \caption{The chemical
	composition of the Earth's crust from
	Ref.~\cite{Emken:2018run}.}\label{tab:earth_crust}
\end{table}

\subsection{Monte Carlo simulation}\label{sub:monte_carlo_simulation}
In the analytic method discussed above, the changes in the direction
of CRDM particles after each scattering process and the possibility of
multiple scatterings are not taken into account, which could lead to
inaccurate description of the CRDM energy loss process.  This problem
becomes serious for CRDM particles with mass and energy significantly
lower than the mass of the nucleus.  In this case the actual
trajectory of the CRDM particle can be much longer than that estimated
using the straight-line propagation approximation.  A more accurate
estimation on the energy loss can be obtained using numerical Monte
Carlo (MC) simulations.

In recent years, a number of public codes are developed for the Earth
attenuation of non-relativistic halo DM\@. The codes based on analytic
method include \earthshadow{}~\cite{Kavanagh:2016pyr} which takes into
account DM deflection based on single scattering process,
and \verne{}~\cite{Kavanagh:2017cru} which considers DM particles
travelling on straight-line trajectories. Both of the codes take into
account the three-dimensional geometry of the Earth which allows for
the study of diurnal modulation.  The attenuation of CRDM based
on \eq{eq:attenuation} is also implemented in
\darksusy{}~\cite{Bringmann:2018lay} with the nuclear form factor neglected.
The codes based on MC simulation such
as \damatis{}~\cite{Mahdawi:2017cxz} takes a flat Earth model and
tracks the vertical distance of the particles only.  More
sophisticated simulations are implemented in the \damascuscrust{} code
for non-relativistic halo DM~\cite{Emken:2018run} which models the
overburden as planar layers.  In the \damascus{}
code~\cite{Emken:2017qmp} a preliminary reference Earth model is
implemented~\cite{Dziewonski:1981xy} and a realistic three dimensional
halo DM injection is applied.  Recently, the \realearthscatterdm{}
code was developed to investigate the daily modulation feature of the
boosted electrophilic DM~\cite{Chen:2021ifo}, which also considered
the density profile of the Earth and focused on the scattering between
DM and bound state electrons.

Since most of the current packages focus on non-relativistic halo DM,
in this work we develop an alternative simulation
code \darkprop{}~\cite{DarkProp:v0.1} which aims at an unified
analysis framework for the Earth attenuation of both the relativistic
and non-relativistic DM particles.  We follow the approach where the
free-propagation length is sampled using the mean-free-path calculated
from the total cross section $\sigma_{\chi N}$ from
integrating \eq{eq:sigma_chi_nuclei} rather than $\sigma^0_{\chi N}$.
The distribution of the scattering angle after the scattering is
sampled according to the differential cross
section \eq{eq:sigma_chi_nuclei} and appropriately normalized.  The
chemical composition of the crust is also taken is to account.  The
condition for the termination of each DM trajectory is that the DM
particle finally leaves the surface of the Earth, or the kinetic
energy is less than a given threshold which cannot trigger a recoil
signal.\note{(number)} This threshold energy is conservatively set to
be $10^{-5}~\text{GeV}$ for all the DM particle masses considered in
this work.  Note that our approach is different from that in
the \damatis{} code and Ref.~\cite{Cappiello:2019qsw} where the
simulation stops when the particle reaches the depth of the
detector. This approach underestimates the underground flux and will
lead to very conservative estimations of the exclusion region of the
DM-nucleon scattering cross section.  The details of the algorithms
and implementations are summarized in
Appendix~\ref{sec:details_of_the_monte_carlo_simulation}.

In this work, for simplicity, we adopted a homogeneous spherical Earth
model with density and chemical composition the same as that used in
our analytic analysis. Although the densities of the core and the
mantle are larger than that of the outer crust, the difference has
little effect on the exclusion region. A more realistic Earth model
should be relevant for the study of the diurnal effect.  We perform
simulations of the DM propagation inside the Earth to obtain the
underground DM flux in the energy range of interest. This is necessary
as light DM particles can enter and leave a given surface within the
Earth multiple times due to the relatively large deflection angle.

As a cross check, we find that our code well-reproduces the results of
the \damascus{} code for non-relativistic halo DM especially the
shielding effect at large cross section~\cite{Emken:2018run}. Details
on the comparison between the two codes can be found in
Appendix~\ref{sec:comparison_with_the_damascus_crust_code_for_halo_dm}.

\subsection{Underground CRDM flux}\label{sub:underground_crdm_flux}

\begin{figure}[tbp]
	\centering \includegraphics[width=0.49\linewidth]{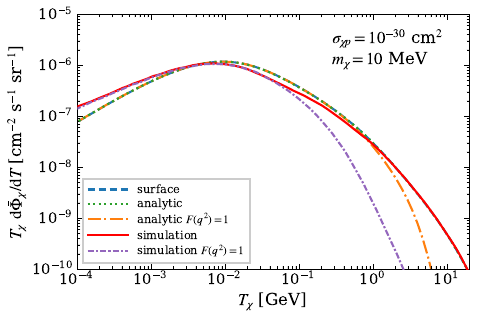} \includegraphics[width=0.49\linewidth]{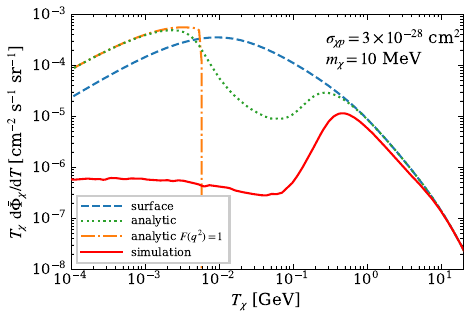} \caption{Underground
	CRDM flux at a depth of 1.4 km as a function of kinetic energy
	with mass $m_\chi = 10~\text{MeV}$ derived from different
	approaches for two cross sections $\sigma_{\chi p} =
	10^{-30}~\text{cm}^2$ (left) and $3\times10^{-28}~\text{cm}^2$
	(right).  The fluxes at the surface of the Earth are indicated
	as blue dashed lines. The green dotted lines (orange
	dash-dotted lines) stand for CRDM fluxes calculated by the
	analytic method with the nuclear form factor considered
	(neglected).  The red solid lines represent the CRDM fluxes
	calculated using the MC simulation with the form factor taken
	into account.  The result of the MC simulation with the form
	factor neglected is not shown in the right panel, as the flux
	is too low due to the strong
	attenuation.}\label{fig:crdm_flux_underground}
\end{figure}

In~\fig{fig:crdm_flux_underground}, we show the underground CRDM
fluxes at a depth of $z = 1.4$~km for a representative DM particle
mass $m_\chi = 10~\text{MeV}$ and cross section $\sigma_{\chi p} =
3\times10^{-28}\text{ cm}^2$ calculated using four different
approaches: analytic approach and MC simulation with and without
including the Helm form factor.
For the analytic method, the attenuated DM flux at depth $z$ is
related to that at surface through the relation
\begin{equation}\label{eq:attenuated_flux}
    \frac{d \Phi}{d T_\chi(z)} = \frac{d \Phi}{d T_\chi(0)} \frac{d
    T_\chi(0)}{d T_\chi(z)} ,
\end{equation}
where the solution of the energy loss equation \eq{eq:attenuation} and
surface CRDM flux obtained in \eq{eq:crdm_flux} are adopted.  For MC
simulation, we simulated $\mathcal{O}(10^7)$ DM particles drawn from
surface CRDM flux with kinetic energy between $10^{-5}$ and
$100~\text{GeV}$. Then the underground flux is produced using the
method detailed in
Appendix~\ref{sec:details_of_the_monte_carlo_simulation}.

\paragraph{Figure 5}
In \fig{fig:crdm_flux_underground}, the underground CRDM fluxes at
depth $z=1.4$~km calculated using different approaches for two typical
cross sections are compared.  The DM particle mass is fixed at
$m_\chi=10~\text{MeV}$. In the left panel, we consider the case with
$\sigma_{\chi p}=10^{-30}~\text{cm}^2$ which corresponds to relatively
small Earth attenuation. In this case, after the form factor is
included, both the analytical approach and MC simulation predict that
the underground CRDM flux is almost the same as that at the
surface. While the methods neglecting the form factor underpredict the
DM flux at high energies above $\sim 1~\text{GeV}$.  In the right
panel, we consider the case with a relatively large cross section of
$\sigma_{\chi p}=3\times 10^{-28}~\text{cm}^2$. The analytic approach
with form factor neglected predicts a sharp cutoff in the CRDM flux at
kinetic energy $6\times 10^{-3}~\text{GeV}$ in accordance
with \eq{eq:Tz_relatv}. Due to this apparent cutoff, the CRDM should
not register a signal at a typical xenon based detector with a
threshold around keV, thus forming a detection ``blind spot'' for DM
detection~\cite{Bringmann:2018cvk,Alvey:2019zaa,Guo:2020drq,Xia:2020apm}. After
the form factor is turned on, the situation changes dramatically. Both
the analytic approach and the MC simulation predict that the CRDM flux
can extend to much higher energy regions and finally approach the
surface flux for DM energy above GeV, which suggests that for such a
large cross section the underground DM can still be energetic enough.

Compared with the analytic approach, the MC simulation predicts
smaller CRDM flux by a few orders of magnitude at low energy below
100~MeV, which is related to the multiple scattering processes during
Earth attenuation.  The ordinary mean-free-path $\ell$ which
characterizes the average free propagation distance between two
successive scatterings (see \eq{eq:mean_free_path} in Appendix) is
actually quite small for the large cross section of
$3\times10^{-28}~\text{cm}^2$ considered in the right panel of Fig.~5.
In this case, it is found to be $\ell\simeq
8.2\times10^{-4}~\text{km}$ for low energy DM particles, which
suggests that the DM particle should go through multiple scatterings
before reaching the underground detector of XENON-1T.  Such
multi-scattering process cannot be well described by the analytical
approach (green dotted). MC simulation (red solid) is more suitable,
as it takes into account the directional change of DM particle in each
scattering.  A consequence of the multi-scattering is that the actual
distance traveled by the DM particles can be much longer than the
depth of the detector $z=1.4$~km. Thus the energy loss is much more
significant than that estimated from the analytical approach.  This
leads to further distortion of the underground DM energy
spectrum. Note that DM particles with lower energies suffer from
larger energy losses as the corresponding cross sections are less
form-factor suppressed.  As it can be seen from the right panel
of \fig{fig:crdm_flux_underground}, compared with the result from the
analytic approach, the numerical MC simulation show more significant
distortion in the lower energy region.  The underground CRDM particles
with $T_\chi \lesssim 0.1$~GeV come dominantly from the energy-loss of
the DM particles in the narrow energy range $0.1-0.5$~GeV on the
surface, which roughly follows a power-law $d\Phi_\chi/d
T_\chi \propto T_\chi^{-2}$.  Due to the spectrum distortion the
underground DM flux changes to roughly $d\Phi_\chi/d T_\chi \propto
T_\chi^{-1}$ which looks approximately like a flat distribution after
rescaled by $T_\chi$ in \fig{fig:crdm_flux_underground}.

In summary, these results show that the effect of the nuclear form
factor is significant in the Earth attenuation. It suppresses the
attenuation of high-energy CRDM regardless of whether the analytic
method or the MC simulation is used.  In addition, the MC simulation
always leads to a stronger attenuation than the analytic approximation
for the mass range under consideration.

\section{Constraints from Direct Detection Experiment}\label{sec:constraints_from_direct_detection_experiment}

\subsection{Recoil spectrum}\label{sub:recoil_spectrum}
Using the underground CRDM flux at depth $z$, the differential nuclear
recoil event rate per target nucleus mass in the direct detection
experiments can be calculated as
\begin{equation}
    \label{eq:recoil_spectrum} \Gamma_N
    = \frac{4\pi}{m_N}\int_{T_\chi^{\min}}^\infty \frac{d \sigma_{\chi
    N}}{d T_N} \frac{d \bar\Phi_{\chi}}{d T_\chi} d T_\chi ,
\end{equation}
where $m_N$ is the mass of the target nucleus $N$ in the detector. In
this work, we focus on the XENON1T experiment, which is a xenon
dual-phase time projection chamber. For XENON1T, $m_N\simeq
122~\text{GeV}$. The form factor in the differential cross section
takes the same form as that the calculation of Earth attenuation.

\begin{figure}[tbp]
	\centering \includegraphics[width=0.7\linewidth]{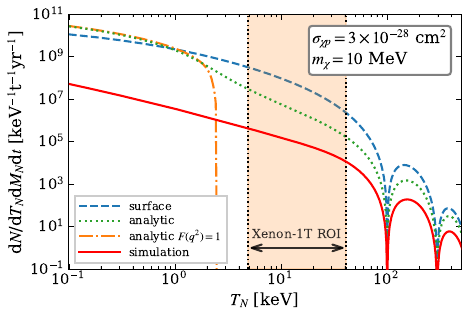} \caption{Recoil
	event rates of the xenon target nuclei from elastic scattering
	with CRDM particles at surface and underground corresponding
	to the CRDM flux in the right panel
	of~\fig{fig:crdm_flux_underground} The region-of-interest
	(ROI) $4.9-40.9~\text{keVnr}$ considered by the XENON1T S1-S2
	data analysis is indicated with vertical black dotted
	lines.}\label{fig:recoil_spectrum}
\end{figure}

\paragraph{Figure 6}
In \fig{fig:recoil_spectrum}, the recoil event spectrum for the large
cross section case corresponds to the same parameters in the right
panel of \fig{fig:crdm_flux_underground} are shown. The results show
that with the form factor included, both the analytic approach and the
MC simulation predict signals in the XENON1T S1-S2 signal region of
interest (ROI) of $4.9-49~\text{keVnr}$. The rapid decreasing and
oscillation of the event rate above the ROI are due to the properties
of the Helm form factor.

\subsection{Exclusion regions}\label{sub:exclusion_region_from_xeon1t_s1_s2_signal}
Since the nuclear recoil event rate from the collisions with CRDM is
quite different from that with the non-relativistic halo DM, to be
more accurate on the effect of threshold, we derive the limits
directly from the distribution of the signals, rather than naively
rescaling the reported experimental limits from halo WIMP
searches~\citep{Bringmann:2018cvk,Dent:2019krz,Bondarenko:2019vrb}.
The theoretical recoil event rate needs to be converted into signals
according to the signal response model of the specific experiment
before it can be compared with the experimental data.  Our analysis is
base on the S1-S2 data of the XENON1T
experiment~\cite{XENON:2018voc}. The XENON1T experiment utilizes the
liquid xenon time projection chambers to detect the recoil energy of
the target nuclei from their scattering with DM particles. The
deposited energy can produce a prompt scintillation signal (S1) and
ionization electrons which are extracted into gaseous xenon and
produce proportional scintillation light (S2). The signal size ratio
of S2/S1 allows for discrimination between nuclear and electronic
recoils.  The S1 and S2 signals will be corrected for their spatial
dependence, which result in the corrected signals of $cS1$ and $cS2_b$
(corresponding to the S2 signals from the bottom photomultiplier
tubes). The detailed signal response model can be found in
Ref.~\cite{XENON:2019izt}.  We numerically calculate the signal
distributions of the scattering between CRDM particles and target
xenon nuclei, and derive the excluded regions in
$(m_{\chi},\sigma_{\chi p})$ plane for the XENON1T data
(S1-S2)~\citep{Aprile:2018dbl,Aprile:2019xxb} using the binned Poisson
statistic approach~\citep{Green:2001xy,Savage:2008er}.  The
distribution of the background events are taken from the XENON1T
analysis.  The calculation procedure of the binned Poisson statistic
approach is the same as that in our previous work~\cite{Xia:2020apm}.

\begin{figure}[tbp]
	\centering \includegraphics[width=0.7\linewidth]{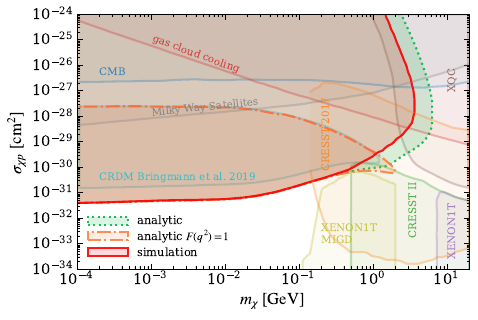} \caption{Exclusion
	regions in the ($m_\chi$, $\sigma_{\chi p}$) plane at $90\%$
	C.L. for CRDM derived from the XENON1T S1-S2 data.  The red
	solid contour represent the exclusion region calculated by the
	MC simulation with the nuclear form factor considered.  The
	result using the analytic method with the form factor
	considered (neglected) is shown with green dotted (orange
	dash-dotted) contours.  The lower bounds of these three
	contours overlap for DM particle mass $m_\chi \lesssim
	0.2~\text{GeV}$ due to the negligible Earth attenuation in
	this region.  A selection of constraints from other direct
	detection
	experiments~\cite{CRESST:2015txj,CRESST:2017ues,XENON:2017vdw,
	XENON:2019zpr} with Earth attenuation
	considered~\cite{Emken:2018run}, XQC~\cite{Mahdawi:2018euy},
	CMB~\cite{Xu:2018efh}, gas cloud
	cooling~\cite{Bhoonah:2018wmw}, and Milky Way
	satellite~\cite{DES:2020fxi} are shown for comparison.  The
	exclusion region derived with a constant effective distance
	$D_\text{eff} = 1~\text{kpc}$ in Ref.~\cite{Bringmann:2018cvk}
	is also shown.}\label{fig:exclusion_region}
\end{figure}

\paragraph{Figure 7}
The exclusion regions at $90\%$ C.L. calculated by different methods
are compared in~\fig{fig:exclusion_region}.  For each DM particle
mass, we scan the values of the scattering cross section to find the
boundaries of the exclusion region.  We find that the exclusion lower
bound reaches~$\sim4\times10^{-32}~\text{cm}^2$ for MeV scale DM,
which is a factor of five improvement compared with that derived under
the assumption of $D_\text{eff}=1$~kpc~\cite{Bringmann:2018cvk}.  For
the cross sections below the exclusion lower bounds, although the
downward CRDM particles reaching the depth of the detector do not
suffer from much energy loss, the upward CRDM particles entering from
the opposite side of the Earth can still loss considerable amount of
energy. If the anisotropy of the CRDM flux in the arrival direction is
considered, there should be a daily modulation of the underground DM
flux, which can be used to explore CRDM in the
future~\cite{Ge:2020yuf}.

In the case of neglecting the nuclear form factor, the exclusion upper
bound obtained from the analytic approach is quite low~$\sim 2\times
10^{-28}~\text{cm}^2$, which is as expected.  After the form factor is
consistently included, we find that the exclusion upper bounds can be
increased at least by four orders of magnitude.  The bounds from the
MC simulation is again more conservative than that from the analytic
approximation.  This new exclusion region, derived from MC simulation
with nuclear form factor considered, is complementary to other
constraints derived from astrophysical and cosmological observables.
It also overlaps with the exclusion regions reported from DM direct
detection experiments with low energy thresholds, especially the
CRESST 2017 surface run. Those limits however, is affected by the
assumption of the DM halo velocity distribution. Especially the value
of the DM escape velocity.  On the other hand, the halo velocity
distribution has basically no effect on the detection of CRDM
particles.

\section{Conclusions}\label{sec:conclusions}

In summary, we have performed a refined analysis on the exclusion
bounds of the DM-nucleon scattering cross section $\sigma_{\chi p}$
using CR boosted DM\@.  We have determined the effective distance
$D_\text{eff}$ using spatial-dependent CR fluxes and included the
contributions from all the major CR nuclear species. We obtained
$D_\text{eff}\simeq 9~\text{kpc}$ for CRDM particle with kinetic
energy above $\sim 1~\text{GeV}$, and found that the corresponding
exclusion lower bounds can be as low as $\sigma_{\chi p} \sim 4\times
10^{-32}~\text{cm}^2$ for DM particle mass around MeV scale and below.
We have simulated the effect of Earth attenuation with the nuclear
form factor fully taken into account.  Using both the analytic and
numerical approaches, we have shown that for the CRDM particles with
mass below GeV scale, the presence of the nuclear form factor strongly
suppress the effect of Earth attenuation.  Consequently, the cross
section that can be excluded by the XENON1T experiment can be many
orders of magnitude higher, which closes the gap in the cross sections
excluded by the XENON1T experiment and that by the astrophysical
observables such as the cosmic microwave background (CMB) and galactic
gas cloud cooling, etc..  In this work we have considered
energy-independent cross sections. It is straightforward to extend the
current framework to energy-dependent cross sections.  Further
improvements that can be made is to consider the anisotropy of CRDM
and the diurnal modulation effect in the future.

\acknowledgments This work is supported in part by
the National Key R\&D Program of China No.~2017YFA0402204, the Key
Research Program of the Chinese Academy of Sciences (CAS), Grant
NO.~XDPB15, the CAS Project for Young Scientists in Basic Research
YSBR-006, and the National Natural Science Foundation of China (NSFC)
No.~11825506, No.~11821505, No.~12047503.

\appendix
\section{Comparison of Dark Matter Profiles}\label{sec:comparison_of_dark_matter_profiles}
In this section, we consider the uncertainties in the CRDM flux from
different DM density profiles. In particular, we compare the effective
distances $D_\text{eff}(T_\chi)$ defined in \eq{eq:deff} predicted
with different DM profiles, as the flux of CRDM is directly
proportional to this quantity.

\begin{figure}[tbp]
    \centering \includegraphics[width=0.49\linewidth]{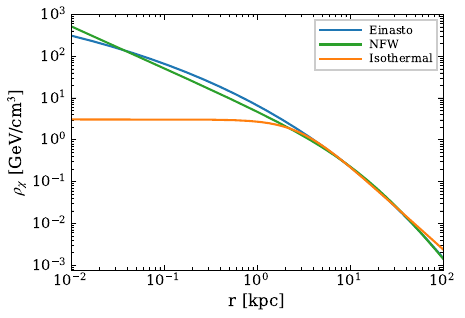} \includegraphics[width=0.49\linewidth]{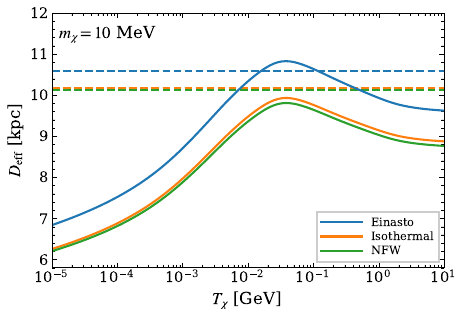} \caption{(Left)
    Einasto, NFW, and Isothermal DM density profiles as a function of
    the radial distance from the Galactic Center.  (Right) Effective
    distance of CRDM derived from different DM density profiles of
    Einasto, Isothermal, and NFW (solid lines).  The DM particle mass
    is fixed at $m_\chi = 10~\text{MeV}$, and the CR fluxes are the
    same as that used in~\fig{fig:d_factor}.  The horizontal dashed
    lines represent the corresponding $D_{\text{eff}}$ assuming a
    homogeneous CR distribution.}\label{fig:deff_dmprofile}
\end{figure}

In addition to the NFW profile in~\eq{eq:nfw} considered in the main
text, we consider the Einasto profile~\cite{Einasto:2009zd} which is a
typical ``Cuspy'' profile, and the Isothermal
profile~\cite{Bahcall:1980fb} which is ``Cored'', for comparison
purpose.  The Einasto profile is parameterised
as~\cite{Einasto:2009zd}
\begin{equation}
    \label{eq:einasto} \rho^\mathrm{EIN}_\chi(r)
    = \rho_\chi^\text{loc} \exp\left[
    - \left(\frac{2}{\alpha}\right) \left(\frac{r^\alpha -
    R_{\odot}^\alpha}{R_\text{s}^\alpha}\right) \right],
\end{equation}
where $\rho^\text{loc}_\chi = 0.3~\text{GeV}/\text{cm}^3$, $\alpha =
0.17$, $R_{\odot} = 8.5~\text{kpc}$, and $R_{\text{s}} =
20~\text{kpc}$.  The Isothermal profile is parameterised
as~\cite{Ackermann:2012rg}
\begin{equation}
    \label{eq:isothermal} \rho_\chi^\text{ISO}(r)
    = \rho_\chi^\text{loc} \frac{R_{\odot}^2 + R_\mathrm{c}^2}{r^2 +
    R_\mathrm{c}^2},
\end{equation}
where $R_\mathrm{c} = 2.8$~kpc. These DM density profiles are shown in
the left panel of~\fig{fig:deff_dmprofile}.

For a given DM particle mass $m_\chi = 10~\text{MeV}$, we repeat the
calculations in~\Sec{sub:calculation_of_the_effective_distance} using
different DM density profiles.  The resulting effective distances are
shown in the right panel of~\fig{fig:deff_dmprofile}. We find that the
Einasto profile leads to the largest $D_{\text{eff}}$, while the NFW
profile leads the most conservative value, and the variation is within
$\sim10\%$. Therefore the difference in the DM density profile near
the Galactic Center has little effect on the effective distance.  If
the CRs are approximated to be uniformly distributed in the CR
propagation halo, the effective distance will be a constant as
discussed in~\eq{eq:deff_const}. These energy-independent effective
distances for different DM density profiles are also shown in the
right panel of~\fig{fig:deff_dmprofile}. The values in this case are
typically larger than that derived from spatial-dependent CR
distributions, because the intensity of the inhomogeneous CRs decrease
as the distance to the galactic plane increases.

\section{Parameters of the Primary CR Sources}\label{sec:parameters_of_the_primary_cr_sources}The primary source spectral parameters of the CR diffusion equation used in
this work are listed
    in~\tab{tab:inj_parameters}.\begin{table}[tbp] \centering \begin{tabular}{clllll}
\toprule
        Nucleus & $\gamma_0^{R_0\text{(GV)}}s_0$ &
                $\gamma_1^{R_1\text{(GV)}}s_1$ &
                $\gamma_2^{R_2\text{(GV)}}s_2$ &
                $\gamma_3^{R_3\text{(TV)}}s_3$ &
                $\gamma_4$ \\ \toprule $_{1}$H & $2.24\, {}^{0.95}\,
                0.29$ & $1.70\, {}^{6.97}\, 0.22$ & $2.44\, {}^{400}\,
                0.09$ & $2.19\, {}^{16}\, 0.09$ & 2.37\\ $_{2}$He &
                $2.05\, {}^{1.00}\, 0.26$ & $1.76\, {}^{7.49}\, 0.33$
                & $2.41\, {}^{340}\, 0.13$ & $2.12\, {}^{30}\, 0.10$ &
                2.37\\ $^{7}_3$Li & \ldots\, ${}^{\ \ \ }$ \ldots &
                $1.10\, {}^{12.0}\, 0.16$ & $2.72\, {}^{355}\, 0.13$ &
                $1.90\, {}^{\ \ }$ \ldots & \ldots\\ $_{6}$C & $1.00\,
                {}^{1.10}\, 0.19$ & $1.98\, {}^{6.54}\, 0.31$ &
                $2.43\, {}^{348}\, 0.17$ & $2.12\, {}^{\ \ }$ \ldots
                & \ldots\\ $^{14}_{7}$N & $1.00\, {}^{1.30}\, 0.17$ &
                $1.96\, {}^{7.00}\, 0.20$ & $2.46\, {}^{300}\, 0.17$ &
                $1.90\, {}^{\ \ }$ \ldots & \ldots\\ $_{8}$O & $0.95\,
                {}^{0.90}\, 0.18$ & $1.99\, {}^{7.50}\, 0.30$ &
                $2.46\, {}^{365}\, 0.17$ & $2.13\, {}^{\ \ }$ \ldots
                & \ldots\\ $_{9}^{19}$F & $0.10\, {}^{1.00}\, 0.30$ &
                $1.00\, {}^{3.80}\, 0.32$ & $3.80\, {}^{8.0}\, 0.35$ &
                $3.12\, {}^{\ \ }$ \ldots & \ldots\\ $_{10}$Ne &
                $0.60\, {}^{1.15}\, 0.17$ & $1.92\, {}^{9.42}\, 0.26$
                & $2.44\, {}^{355}\, 0.17$ & $1.97\, {}^{\ \ }$ \ldots
                & \ldots\\ $_{11}$Na & $0.50\, {}^{0.75}\, 0.17$ &
                $1.98\, {}^{7.00}\, 0.21$ & $2.49\, {}^{355}\, 0.17$ &
                $2.14\, {}^{\ \ }$ \ldots & \ldots\\ $_{12}$Mg &
                $0.20\, {}^{0.85}\, 0.12$ & $1.99\, {}^{7.00}\, 0.23$
                & $2.48\, {}^{355}\, 0.17$ & $2.15\, {}^{\ \ }$ \ldots
                & \ldots\\ $_{13}$Al & $0.20\, {}^{0.60}\, 0.17$ &
                $2.04\, {}^{7.00}\, 0.20$ & $2.48\, {}^{355}\, 0.17$ &
                $2.14\, {}^{\ \ }$ \ldots & \ldots\\ $_{14}$Si &
                $0.20\, {}^{0.85}\, 0.17$ & $1.97\, {}^{7.00}\, 0.26$
                & $2.47\, {}^{355}\, 0.17$ & $2.19\, {}^{\ \ }$ \ldots
                & \ldots\\ $_{15}$P & $0.25\, {}^{1.60}\, 0.19$ &
                $1.95\, {}^{7.00}\, 0.20$ & $2.48\, {}^{355}\, 0.17$ &
                $2.14\, {}^{\ \ }$ \ldots & \ldots\\ $_{16}$S &
                $0.80\, {}^{1.30}\, 0.17$ & $1.96\, {}^{7.00}\, 0.20$
                & $2.49\, {}^{355}\, 0.17$ & $2.14\, {}^{\ \ }$ \ldots
                & \ldots\\ $_{17}$Cl & $1.10\, {}^{1.50}\, 0.17$ &
                $1.98\, {}^{7.20}\, 0.20$ & $2.53\, {}^{355}\, 0.17$ &
                $2.14\, {}^{\ \ }$ \ldots & \ldots\\ $_{18}$Ar &
                $0.20\, {}^{1.30}\, 0.17$ & $1.96\, {}^{7.00}\, 0.20$
                & $2.46\, {}^{355}\, 0.17$ & $2.09\, {}^{\ \ }$ \ldots
                & \ldots\\ $_{19}$K & $0.20\, {}^{1.40}\, 0.15$ &
                $1.96\, {}^{7.00}\, 0.20$ & $2.53\, {}^{355}\, 0.17$ &
                $2.14\, {}^{\ \ }$ \ldots & \ldots\\ $_{20}$Ca &
                $0.30\, {}^{1.00}\, 0.11$ & $2.07\, {}^{7.00}\, 0.20$
                & $2.48\, {}^{355}\, 0.17$ & $2.14\, {}^{\ \ }$ \ldots
                & \ldots\\ $_{21}$Sc & $0.20\, {}^{1.40}\, 0.17$ &
                $1.97\, {}^{7.00}\, 0.22$ & $2.53\, {}^{355}\, 0.17$ &
                $2.14\, {}^{\ \ }$ \ldots & \ldots\\ $_{22}$Ti &
                $1.50\, {}^{0.90}\, 0.17$ & $1.98\, {}^{7.00}\, 0.22$
                & $2.57\, {}^{355}\, 0.17$ & $2.14\, {}^{\ \ }$ \ldots
                & \ldots\\ $_{23}$V & $1.10\, {}^{0.80}\, 0.17$ &
                $1.98\, {}^{7.00}\, 0.22$ & $2.53\, {}^{355}\, 0.17$ &
                $2.14\, {}^{\ \ }$ \ldots & \ldots\\ $_{24}$Cr &
                $1.70\, {}^{0.65}\, 0.17$ & $1.99\, {}^{7.00}\, 0.20$
                & $2.48\, {}^{355}\, 0.17$ & $2.14\, {}^{\ \ }$ \ldots
                & \ldots\\ $_{25}$Mn & $0.20\, {}^{0.85}\, 0.10$ &
                $2.08\, {}^{7.00}\, 0.20$ & $2.48\, {}^{355}\, 0.17$ &
                $2.14\, {}^{\ \ }$ \ldots & \ldots\\ $_{26}$Fe &
                $0.95\, {}^{2.00}\, 0.20$ & $3.62\, {}^{2.94}\, 0.10$
                & $2.05\, {}^{17.0}\, 0.18$ & $2.452\, {}^{0.355}\,
                0.17$ & $2.23$ \\ $_{27}$Co & $0.80\, {}^{0.70}\,
                0.15$ & $1.98\, {}^{7.00}\, 0.20$ & $2.49\, {}^{355}\,
                0.17$ & $2.14\, {}^{\ \ }$ \ldots & \ldots\\ $_{28}$Ni
                & $1.50\, {}^{0.65}\, 0.17$ & $1.98\, {}^{7.00}\,
                0.20$ & $2.48\, {}^{355}\, 0.17$ & $2.14\, {}^{\ \
                }$ \ldots
                & \ldots\\ \botrule \end{tabular} \caption{The primary
                source spectral parameters of the CR diffusion
                equation taken from Ref.~\cite{Boschini:2020jty} with
                updated parameters of Fe from
                Ref.~\cite{Boschini:2021gdo} and F from
                Ref.~\cite{Boschini:2021ekl}.}\label{tab:inj_parameters}
\end{table}

\section{Details of the Monte Carlo Simulation}\label{sec:details_of_the_monte_carlo_simulation}
In this section we describe the details of the MC simulation in this
work including the initial conditions, the trajectory simulation, and
the reconstruction of the underground CRDM flux.

\subsection{Initial conditions}\label{sub:initial_conditions}
In this work, we only consider the angular-averaged CRDM flux as most
of the current DM direct detection experiments cannot distinguish the
arrival direction of DM\@.  The initial momenta of the particles are
set to be isotropic with the kinetic energy sampled from the
normalized surface flux in~\eq{eq:crdm_flux_k_factor}.  The isotropy
of the injection is guaranteed by appropriately choosing the initial
position~\cite{Emken:2017qmp}. Once a random direction is sampled
isotropically, the initial position is chosen randomly on a circular
disk of radius $R_\oplus$ at a distance $R$ from the Earth center and
perpendicular to that direction. Here $R_\oplus \approx
6371~\text{km}$ is the radius of the Earth, and $R = 1.1 R_\oplus$ is
an arbitrary distance greater than $R_\oplus$. The momentum is then
taken to be along that direction and point to the Earth.  We have
checked that this approach ensures that the same isotropic flux is
obtained inside the Earth in the case where the Earth is transparent
to DM particles.

\subsection{Trajectory simulation}\label{sub:trajectory_simulation}
The simulation of the trajectory of DM particles inside the Earth is
composed of two processes, free-propagation and scattering. The
free-propagations are considered as straight lines of length $\hat{l}$
between two consecutive collisions since gravity can be safely
neglected. The cumulative distribution function (CDF) of the random
variable $\hat{l}$ is
\begin{equation}
    F(L) = P(\hat{l} < L) = 1 - \exp\left(-\int_0^L \frac{d
    x}{\ell}\right), \label{eq:CDF_l}
\end{equation}
where the mean free path $\ell$ is defined as
\begin{equation}\ell^{-1} = \sum_N \ell_N^{-1}
    \equiv \sum_N n_N \sigma_{\chi N}, \label{eq:mean_free_path}
\end{equation}
where $n_N$ is the number density of nuclear species $N$, and
$\sigma_{\chi N}$ is the total cross section defined
in~\eq{eq:sigma_ratio}.  We draw samples of the free path by using the
inverse CDF method
\begin{equation}
    L = F^{-1}(\xi), \label{eq:draw_l}
\end{equation}
where $\xi$ is a random number drawn from a uniform distribution
between 0 and 1. Explicitly, we solve $L$ from the equation
\begin{equation}
    \int_0^L \frac{d x}{\ell} = -\ln(1-\xi).  \label{eq:free_path}
\end{equation}
The specific form of $\ell$ depends on the cross section and the Earth
model. For the homogeneous Earth model considered in this work, $\ell$
is independent of position, and the free path is solved as
\begin{equation}
    L = -\ell \ln(1-\xi).  \label{eq:free_path_homogeneous}
\end{equation}

After traveling freely for a distance of $L$, the DM particle scatters
with the nuclei in the Earth. Which kind of nucleus to collide with is
determined by the collision probability
\begin{equation}
    P_N = \frac{n_N \sigma_{\chi N}} {\sum_{N'}n_{N'} \sigma_{\chi
    N'}} = \frac{\ell}{\ell_N}.  \label{eq:collision_probability}
\end{equation}
Then we simulate the scattering between DM and the nucleus $N$
according to the differential cross section. The probability density
function (PDF) of the momentum transfer squared $q^2$ is
\begin{equation}
    p(q^2) = \frac{1}{\sigma_{\chi N}} \frac{d \sigma_{\chi N}}{d
    q^{2}}.
\end{equation}
For the cross section considered in this work
in~\eq{eq:sigma_chi_nuclei}, sampling via inversion of the CDF
simplifies to solve $q^2$ from the equation
\begin{equation}
    \frac{\int_0^{q^2} F^2_N(q'^2)d q'^2} {\int_0^{q^2_{\max}}
    F^2_N(q'^2)d q'^2} = \xi, \label{eq:sampling_q2}
\end{equation}
where $q^2_{\max} = 2m_N T_N^{\max}$ with $T_N^{\max}$
from~\eq{eq:kin_tb_max}.  After the recoil energy $T_N = q^2 / 2m_N$
was sampled, the scattering angle of DM particle in the laboratory
frame can be calculated from kinematics as
\begin{equation}
    \cos\theta = \frac{p_i^2 - T_N (T_i + m_\chi + m_N)}{p_i
    p_f}, \label{eq:scattering_angle}
\end{equation}
where $p_{i(f)} = \sqrt{T_{i(f)}^{2} + 2m_\chi T_{i(f)}}$ is the
initial (final) momentum of DM with $T_f = T_i - T_N$.  The direction
of the final momentum is determined by supplementing a uniformly
random azimuthal angle from $0$ to $2\pi$.

\subsection{Underground flux reconstruction}\label{sub:underground_flux_reconstruction}
With the continuous injection of DM particles, we expect them to form
a stable distribution inside the earth. But simulating the time
evolution is time-consuming because the simulated trajectories before
reaching equilibrium are all discarded. On the other hand, all the
trajectories can be regarded as an ensemble to reconstruct the
underground flux.  We record the crossing events of particles passing
through a given surface $\Delta S$ underground to calculate the
corresponding flux. An effective time interval $\Delta t$ can be
determined by the injected particle number and the flux. Assuming time
translation invariance, the collected events in the simulation are
equivalent to those in the time interval $\Delta t$ from a continuous
injection scenario. Then the underground flux is related to these
crossing events as
\begin{equation}
    \label{eq:flux_simulation_def} \frac{d \Phi_\chi^d}{d T_\chi}
    |\cos\theta_i| = \frac{\Delta N_i}{\Delta T_\chi \Delta
    S \Delta\Omega \Delta t},
\end{equation}
where $\Delta N_i$ is the number of events crossing the area $\Delta
S$ with energy in the $i$th energy bin $(T_i, T_i + \Delta T_\chi)$
and from a solid angle $\Delta\Omega$ of the arrival
direction. $\theta_i$ is the angle of the arrival direction.

We sample $N_\text{sim}$ DM particles using the method described in
Appendix~\ref{sub:initial_conditions} with energies in an interval
$(T_a, T_b)$.  We trace each particle during the simulation and record
the crossing events whenever it passes through an underground
spherical surface at a depth of $d$ centered at the center of the
Earth.  Each crossing event contains the kinetic energy, position, and
momentum direction of the particle.  In this case, we have
$\Delta\Omega = 4\pi$ and $\Delta S = S_d = 4\pi {(R_\oplus - d)}^2$,
which is the area of the underground sphere.  The effective injection
time $\Delta t$ is determined as
\begin{equation}
    \label{eq:injected_particle_number} \Delta t
    = \frac{N_{\text{sim}}}{\pi S_0 \Phi_0^\text{norm}},
\end{equation}
where $\Phi_0^\text{norm}=\int_{T_a}^{T_b}\frac{d\Phi_{\chi}^0} {d
T_\chi}d T_\chi$ is the integral flux of CRDM at the surface of the
Earth, and $S_0 = 4\pi R_\oplus^2$ is the Earth's surface area. Note
that this expression is derived using the assumption that the flux is
isotropic and we only inject particles from outside the Earth.
Combining these terms we can write the angular-averaged underground
flux as
\begin{equation}
    \label{eq:flux_simulation_calc} \frac{d \bar\Phi^d_\chi} {d
    T_\chi} = \frac{S_0\Phi_0^\text{norm}} {4 S_d N_\text{sim}} \sum_j
    w_j f(T),
\end{equation}
where $w_j = 1 / |\cos\theta_j|$ is the weight of event $j$, and
$f(T)$ is the normalized probability density function of kinetic
energy constructed from the sample points weighted by $w_j$.  If
importance sampling (IS) is implemented in the simulation, the
corresponding weights need to be transferred to $w_j$.  We adopted IS
to improve the sampling efficiency of the high-energy part of the
injection flux, which is summarized in
Appendix~\ref{sec:importance_sampling}.

Note that the formula of Eq.~\eqref{eq:flux_simulation_calc} is valid
under the condition that the initial energy of the particles is
sampled from the flux, which is essentially different from sampling
from velocity distribution.  But if we regard sampling from the
velocity distribution as an IS from the flux, the above analysis can
be followed by acquiring a weight for each DM particle. The weight is
proportional to the initial speed $v$ since the flux $\Phi \propto v
f(\bm{v})$.  The \damascus{} code samples initial state from the
velocity distribution with this weight missed in the original
version~\cite{Emken:2017qmp}, which has been corrected in the latest
version~\footnote{\url{https://github.com/temken/DaMaSCUS}}.  Our
flux-based method is consistent with the updated version.

\section{Comparison with the \damascuscrust{} Code for Halo DM}\label{sec:comparison_with_the_damascus_crust_code_for_halo_dm}
In this section we give the results of the Earth attenuation of the
non-relativistic halo DM and compare them with the results from the
\damascuscrust{}~\cite{Emken:2018run}. We show that the results are in good
agreement. We further show that the attenuation in the MC simulation
can be stronger than that in the analytic method for light halo DM\@.
Our code is publicly available at the \darkprop{}
webpage~\cite{DarkProp:v0.1}.

The simulation procedure is essentially the same as that for CRDM
described in Appendix~\ref{sec:details_of_the_monte_carlo_simulation},
except that the initial condition is based on the Maxwellian velocity
distribution of the standard halo model (SHM). The most probable speed
and the escape velocity are taken to be $v_0 = 220~\text{km}\text{s}$
and $v_\text{esc} = 544~\text{km}/\text{s}$ respectively.  The initial
velocity is sampled from the isotropic SHM velocity distribution and
then subtracted by a fixed oriented Earth velocity with a magnitude of
$240~\text{km}/\text{s}$.  We perform two sets of simulations for DM
particle mass $m_\chi = 10$ and $0.5~\text{GeV}$. Each set of the
simulation involve a number of cross sections.  The underground DM
fluxes at a depth of $1.4~\text{km}$ are reconstructed.  We calculate
the recoil events number of the DM-xenon nucleus scatterings in the
recoil energy range of $0.5 - 40~(0.013 - 40)~\text{keV}$ for DM
particle mass $m_\chi = 10~(0.5)~\text{GeV}$ with a exposure of
$35.6~\text{ton}\cdot\text{days}$.  The recoil energy ranges are
chosen to be different and adjusted for producing the same events
numbers in two DM particle mass cases when Earth attenuation is
absent. The threshold velocity for termination of the simulation is
set to be $566~(536)~\text{km}/\text{s}$ for $m_\chi =
10~(0.5)~\text{GeV}$.  The nuclear form factor is neglected for these
light non-relativistic DMs.

\begin{figure}[tbp]
    \centering \includegraphics[width=0.7\linewidth]{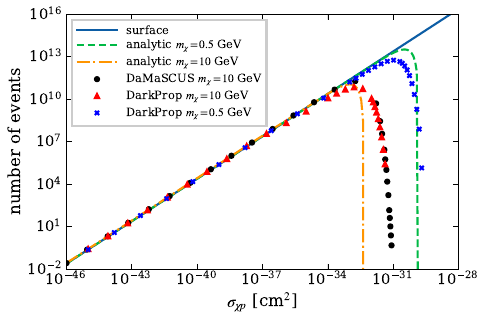} \caption{Recoil
    events number predicted for a xenon based DM detector as a
    function of cross section for halo DM with masses $m_\chi = 10$
    and $0.5~\text{GeV}$. The exposure is assumed to be
    $35.6~\text{ton}\cdot\text{days}$, and the considered nuclear
    recoil energy range is $[5, 40]~\text{keV}$ for $m_\chi =
    10~\text{GeV}$ and $[0.013, 40]~\text{keV}$ for $m_\chi =
    0.5~\text{GeV}$ respectively.  The green dashed (orange
    dash-dotted) line represents the events number calculated using
    analytic method for $m_\chi = 10~(0.5)~\text{GeV}$.  The events
    numbers at the surface for the two DM particle masses are the same
    due to the specific recoil energy ranges chosen and shown with a
    blue solid line.  The result using MC simulation for $m_\chi =
    10~(0.5)~\text{GeV}$ is shown as red triangles (blue crosses). The
    result from~\damascuscrust{}~\cite{Emken:2018run} is shown as
    black circles for comparison.}\label{fig:repeatDaMaSCUS}
\end{figure}

The recoil events numbers derived from the MC method as a function of
cross section are shown in~\fig{fig:repeatDaMaSCUS}, which are
compared with the results using the analytic method. The unattenuated
events number on the surface of the Earth is also shown for
comparison.
In the case of $m_\chi = 10~\text{GeV}$, the number of events drops
rapidly when the cross section is sufficiently large and drops later
in the simulation than that in the analytic approximation which is in
good agreement with \damascuscrust{}~\cite{Emken:2018run}. On the
other hand, for lighter DM with mass $m_\chi = 0.5~\text{GeV}$, the
number of events in the simulation drops before that in the analytic
approximation as the cross section increases, which is opposite to the
case of $m_\chi = 10~\text{GeV}$.  This result shows that the
attenuation in the MC simulation can be stronger than that in the
analytic method for light halo DM due to larger scattering angles in
the laboratory frame.

\section{Importance Sampling}\label{sec:importance_sampling}
The CRDM flux is approximately a power law spectrum with a power index
of about~$-2$~\cite{Xia:2020apm}, therefore dropping quickly with
energy increases.  If we sample energy points directly from the
normalized flux in a wide energy range, the energies at the
high-energy end will be difficult to obtain due to the extremely low
probability.  We use importance sampling method to overcome this
problem since we are concerned about the Earth attenuation of
high-energy CRDM\@.

Suppose $f(T)$ is the probability distribution of energy we need to
sample.  We do not sample directly from this distribution but from an
artificial flatter distribution $g(T)$, in which the relative
probability of the high-energy part is higher. Then we assign a weight
$w_i = f(T_i)/g(T_i)$ to each sample $T_i$.  The original distribution
$f(T)$ should be reconstructable using these weighted samples. In MC
simulation, these weights need to transfer to the final sample
analysis.

In our problem, the initial energy of each trajectory is sampled from
the surface CRDM flux. We choose $g(T)$ as a logarithmic uniform
distribution on the region-of-interest $(T_a, T_b)$
\begin{equation}
    \label{eq:log_uniform_pdf} g(T) = \frac{1}{(\ln T_b - \ln T_a)T}.
\end{equation}
Then the energy points $T_j$ are obtained directly by the exponential
of samples from a uniform distribution on $(\ln T_a, \ln T_b)$ with
the weights
\begin{equation}
    \label{eq:IS_weights} w^\text{IS}_j
    = \frac{T_j}{\Phi_0^\text{norm}} \frac{d\Phi^0_\chi}{d
    T_j} \ln \left(\frac{T_b}{T_a}\right).
\end{equation}
Correspondingly, the weights in Eq.~\eqref{eq:flux_simulation_calc}
for calculating the underground flux need to be modified to
\begin{equation}
    \label{eq:weights_modified} w_j
    = \frac{w^\text{IS}_j}{|\cos\theta_j|}.
\end{equation}
If a particle crosses the underground surface multiple times, these
events will acquire the same weight $w^\text{IS}$ according to the
initial energy.

\bibliographystyle{arxivref}\bibliography{crdm_ref_spire}

\end{document}